%% file: main.tex
\documentclass[twocolumn]{aastex63}
\usepackage{fp}
\usepackage{bm}
\usepackage{xcolor}
\DeclareSymbolFont{matha}{OML}{txmi}{m}{it}
\DeclareMathSymbol{\varv}{\mathord}{matha}{118}
\usepackage{subfigure}

\shorttitle{DES SLSNe}
\shortauthors{Hsu, Hosseinzadeh, \& Berger}

\input{affil.tex}

\begin{document}

\title{Magnetar Models of Superluminous Supernovae from the Dark Energy Survey: Exploring Redshift Evolution}

\correspondingauthor{Brian Hsu}
\email{brianhsu@college.harvard.edu}

\author[0000-0002-9454-1742]{Brian Hsu}
\CfA

\author[0000-0002-0832-2974]{Griffin~Hosseinzadeh}
\CfA

\author[0000-0002-9392-9681]{Edo~Berger}
\CfA

\begin{abstract}
Superluminous supernovae (SLSNe) are luminous transients that can be detected to high redshifts with upcoming optical time-domain surveys such as the Vera C.~Rubin Observatory Legacy Survey of Space and Time (LSST).  An interesting open question is whether the properties of SLSNe evolve through cosmic time.  To address this question, in this paper we model the multi-color light curves of all 21 Type I SLSNe from the Dark Energy Survey (DES) with a magnetar spin-down engine, implemented in the Modular Open-Source Fitter for Transients (\texttt{MOSFiT}). With redshifts up to $z\approx 2$, this sample includes some of the highest-redshift SLSNe. We find that the DES SLSNe span a similar range of ejecta and magnetar engine parameters to previous samples of mostly lower-redshift SLSNe (spin period $P\approx 0.79-13.61$ ms, magnetic field $B\approx (0.03-7.33)\times10^{14}$ G, ejecta mass $M_{\rm ej}\approx 1.54-30.32$ M$_{\odot}$, and ejecta velocity $\varv_{\rm ej}\approx (0.55-1.45)\times 10^4$ km s$^{-1}$). The DES SLSN sample by itself exhibits the previously found negative correlation between $M_{\rm ej}$ and $P$, with a pronounced absence of SLSNe with low ejecta mass and rapid spin. Combining our results for the DES SLSNe with 60 previous SLSNe modeled in the same way, we find no evidence for redshift evolution in any of the key physical parameters. 
\end{abstract}

\section{Introduction} 
\label{sec:intro}

Type I superluminous supernovae (hereafter, SLSNe) are a rare subclass of core-collapse supernovae (CCSNe) that have been discovered by wide-field time-domain optical surveys over the past decade \citep{Chomiuk_2011,Quimby_2011}.
They were originally defined to have a peak absolute magnitude of $M<-21$ \citep{Gal-Yam_2012}, but are now defined spectroscopically by an absence of hydrogen features, blue continua, and, usually, unique early-time ``W''-shaped \ion{O}{2} absorption lines at $\sim 3600-4600$ \AA\ \citep[e.g.][]{Lunnan_2013,Mazzali_2016,Quimby_2018}.

SLSNe radiate 10-100 times more energy in the UV/optical compared to normal CCSNe, and generally exhibit longer durations \citep[e.g.][]{Nicholl_2015b,Inserra_2017,Lunnan_2018,DeCia_2018}. As a result of their light curve behavior and spectral properties, it has become clear that SLSNe are not powered by radioactive decay of $^{56}$Ni as in normal hydrogen-poor CCSNe.  Interaction with a hydrogen-poor circumstellar medium could in principle provide sufficient radiative energy, but there is no clear spectroscopic evidence for such interaction  \citep{Jerkstrand_2017,Liu_2017b}.

Instead, both the diverse light curves and the spectral evolution of SLSNe can be explained well with a central engine, namely a rapidly spinning ($\sim {\rm few}$ ms) and highly magnetized ($\sim 10^{14}$ G) neutron star \citep[a ``magnetar";][]{Kasen_Bildsten_2010,Woosley_2010,Dessart_2012,Metzger_2015,Nicholl_2017b}.  This model accounts for the broad range of peak luminosities and timescales (e.g., \citealt{Nicholl_2017b,Blanchard_2020}), for the early UV/optical spectra (e.g., \citealt{Nicholl_2017c}), for the nebular phase spectra (e.g., \citealt{Nicholl_2016b,Nicholl_2019,Jerkstrand_2017}), and, most recently, for the power law decline rate observed in SN\,2015bn at $\sim 10^3$ d \citep{Nicholl_2018}.  Additional support for a magnetar engine comes from the low metallicity host galaxies of SLSNe, which most closely resemble the hosts of long gamma-ray bursts, another rare population of CCSNe that are powered by a central engine (e.g., \citealt{Lunnan_2014,Perley_2016}, but cf.\ \citealt{Schulze_2018}).

Systematic analyses of some SLSN samples in the context of a magnetar engine model have been carried out in recent years by \citet{Nicholl_2017b}, \citet{Villar_2018}, and \citet{Blanchard_2020} using the semi-analytical Modular Open-Source Fitter for Transients \citep[\texttt{MOSFiT};][]{Guillochon_2018}.  Recently, \citet{Angus_2019} published a sample of 21 SLSNe at $z\approx 0.2-2$ discovered by the Dark Energy Survey (DES), presented their general observed properties, and derived physical properties from magnetar model fits to Gaussian process interpolated pseudo-bolometric light curves.  Here, we use \texttt{MOSFiT} to model the multi-color light curves of these DES SLSNe with a magnetar engine for the first time to systematically compare them to the previously-published SLSN samples.  Taking advantage of the broad redshift range of the DES sample, we also use the combined sample of 81 SLSNe to explore any redshift evolution in the engine and SN ejecta parameters.  

\defcitealias{Planck_2016}{Planck Collaboration 2016}
\defcitealias{Planck_2020}{Planck Collaboration 2020}
The paper is structured as follows. The DES SLSN sample is summarized in \S\ref{sec:data}.  The model fitting procedure is described in \S\ref{sec:model}. The resulting fits and model parameters are presented in \S\ref{sec:results}, while in \S\ref{sec:analysis} we place these results in the context of previous SLSN samples analyzed in the same manner. Throughout the paper, we assume a flat $\Lambda$CDM cosmology with $\Omega_\mathrm{m}=0.308$ and $H_0=67.8\ \text{km}\ \text{s}^{-1}\ \text{Mpc}$, based on the Planck 2015 results \citepalias{Planck_2016}.

\section{Data Set}
\label{sec:data}

We model the DES SLSN sample presented in \citet{Angus_2019}. DES \citep{DES_2005,DES_2016} was an optical imaging survey covering 5000 deg$^2$ of the southern sky, using the 570 megapixel Dark Energy Camera \citep[DECam;][]{Flaugher_2015} on the 4-m Blanco Telescope at Cerro Tololo Inter-American Observatory in Chile. The DES-SN component of the survey \citep{Berstein_2012}, accounting for $\approx 10\%$ of the total survey time, covered ten 3-deg$^2$ individual fields in \textit{g}, \textit{r}, \textit{i}, \textit{z} with a roughly weekly cadence to typical limiting magnitudes of $\approx 23.5$ (8 fields) and $\approx 24.5$ (2 fields). See \citet{Kessler_2015} for a more detailed description of the observing strategy, and \citet{Morganson_2018} for the DES image processing pipeline. 

\begin{figure}[t!]
    \centering
    \includegraphics[width=\columnwidth]{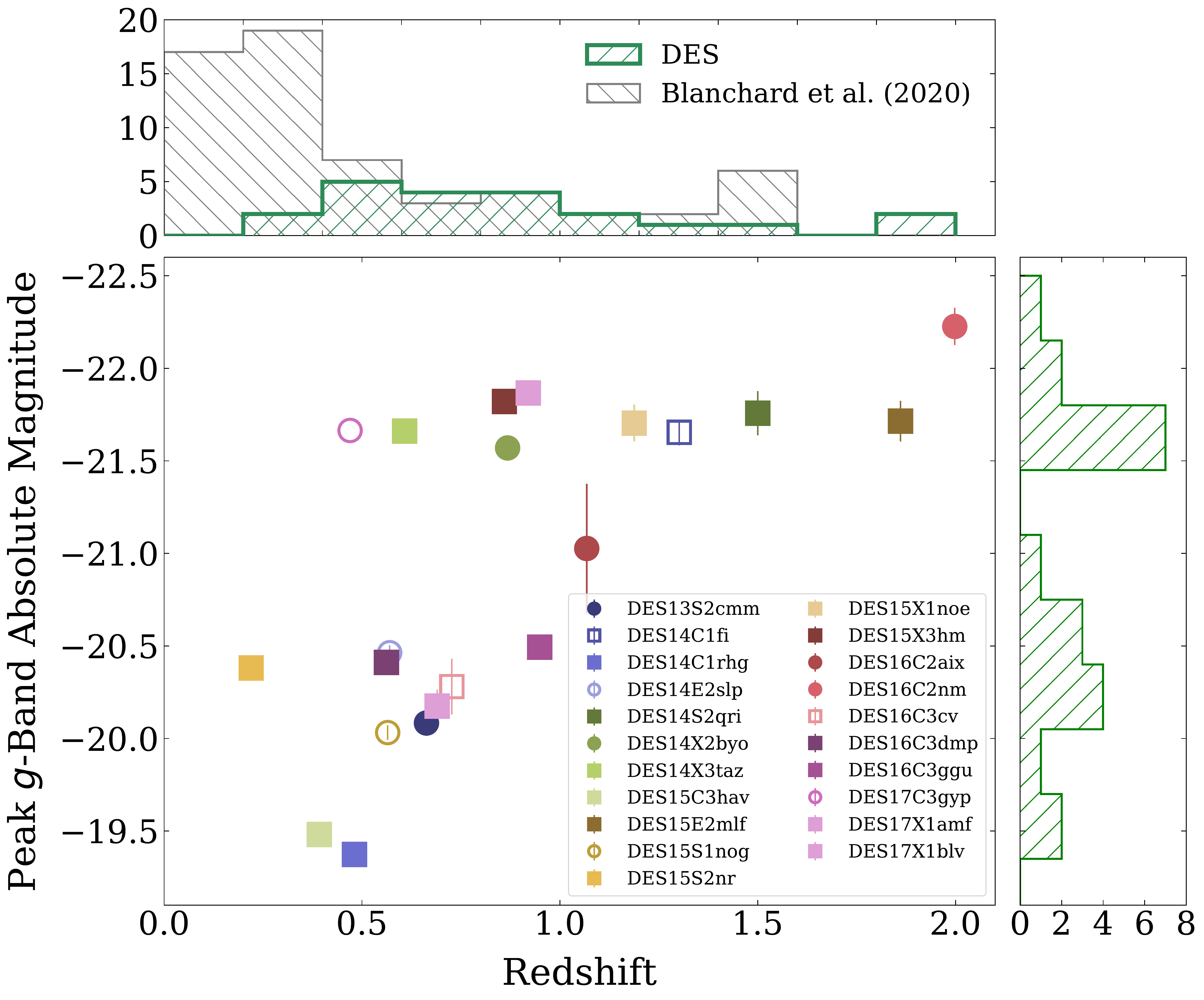}
    \caption{Rest-frame peak $g$-band absolute magnitude versus redshift for the DES SLSNe. The top panel shows the redshift distribution of the DES SLSNe (green) compared to the SLSN compilation recently analyzed by \citet{Blanchard_2020}. Square symbols indicate observed pre-peak bumps in the light curve via visual inspection. Solid and open markers indicate the Gold and Silver spectroscopic classification, respectively, of \citet{Angus_2019}.}
    \label{fig:mag_z}
\end{figure}

\citet{Angus_2019} classified transient events as SLSNe using \texttt{SUPERFIT} \citep{Howell_2005} with the SLSN spectral template library of \citet{Quimby_2018}. Their final sample consists of 21 SLSNe with a redshift range of $z=0.220-1.998$, separated into two classification ``standards'' of Gold (16 events) and Silver (5 events) based on their confidence in the spectroscopic classification, listed in Table~\ref{tab:kcorrection}. In the following analysis we correct the photometric measurements in \citet{Angus_2019} for Milky Way extinction using \citet{Schlafly_Finkbeiner_2011}.

\input{kcorrection}

In Figure~\ref{fig:mag_z} we plot the observed peak absolute magnitudes versus redshift for the DES sample, as well as the redshift distribution in comparison to the SLSN compilation from \citet{Blanchard_2020}. Because we do not have a complete set of spectra for the DES SLSNe, we correct our observed peak magnitudes to a single rest-frame filter by calculating synthetic photometry in the rest-frame $g$-band and the closest observed filter using the modified blackbody SED returned by \texttt{MOSFiT} (see \S \ref{sec:model}) at the epoch corresponding to the observed peak magnitude. See Table~\ref{tab:kcorrection} for a complete list of redshift, observer-frame filter used, $K$ correction, and rest-frame peak $g$-band absolute magnitude for each DES SLSN.

The peak absolute magnitudes span a wide range of $\approx -19.5$ to $\approx -22.5$, with an apparent gap\footnote{There is one event, DES16C2aix, that has a peak rest-frame $g$-band absolute magnitude in this region. However, we note that the peak magnitude for DES16C2aix has large uncertainties and the next brightest point in rest-frame $g$-band resides on the lower bound of the plotted peak magnitude.} at about $-20.5$ to $-21.5$. As expected, the lower luminosity events are mostly confined to lower redshift ($z\lesssim 1$) while the more luminous events extend over $z\approx 0.5-2$. \cite{Angus_2019} suspected that the broad spread in peak magnitudes is due to varying injection times of magnetar power with respect to explosion times, where delayed injections cause systematically fainter peak magnitudes. The apparently bimodal peak magnitude distribution, with roughly equal numbers in each group, indicates that the lower luminosity events are intrinsically more common than the high luminosity events. In Section~\ref{sec:analysis}, we explore whether this bimodality is reflected in the physical parameters of the magnetar engine and/or ejecta.  

In Figure~\ref{fig:fits} we show the full multi-band light curves of the 21 SLSNe.  The light curves are well sampled and exhibit a wide range of morphologies in terms of rise and decline timescales. We also note that some light curves are smooth (e.g., DES15X3hm) while other exhibit eqaul-magnitude double peaks (e.g., DES16C3cv) or pronounced pre-peak ``bumps" (e.g., DES15C3hav). A detailed discussion of the phenomenological light curve properties is provided by \citet{Angus_2019}.

\section{Magnetar Engine Model}
\label{sec:model}

We fit the optical light curves of the 21 DES SLSNe using the Modular Open-Source Fitter for Transients \citep[\texttt{MOSFiT};][]{Guillochon_2018} with the magnetar spin-down model described in \citet{Nicholl_2017b}. In short, \texttt{MOSFiT} is a open-source, \texttt{Python}-based light curve fitting package that employs a Markov chain Monte Carlo (MCMC) algorithm to fit analytical models to multi-band light curves. Typically, blackbody SEDs can reproduce the optical and NIR broadband light curves of SLSNe \citep{Nicholl_2016a,Nicholl_2017a} due to their relatively featureless spectra \citep{Yan_2017}. However, it has been shown that the observed SED is subject to absorption at UV wavelengths \citep{Mazzali_2016,Quimby_2018} that will dominate at observer-frame optical wavelengths of higher redshift SLSNe, such as ones from DES. We therefore assume a modified SED where the overall shape follows a blackbody distribution but linearly suppressed blueward of $\approx 3000$ \AA\ \citep[see Figure 1 in][]{Nicholl_2017b}. \citet{Nicholl_2017b} also showed that this simplified SED assumption applies to a wide range of redshifts, and does not lead to systematic differences between the light-curve fits of higher and lower redshift SLSNe.

The magnetar model has 12 free parameters, of which 8 are nuisance parameters that we marginalize over to obtain the 4 physical parameters related to the ejecta and engine properties. We set one of the nuisance parameters, the angle $\theta_{PB}$ between the magnetic field and the rotational axis of the magnetar, to be constant at $90^{\circ}$. This choice ensures that the derived $B$-field strength is a lower limit \citep[for more details, see][]{Nicholl_2017b}. The nuisance parameters are not well-constrained by the model. Events with sufficient late-time observations, however, are able to constrain the $\gamma$-ray opacity $\kappa_{\gamma}$.  The neutron star mass is degenerate with the spin period and magnetic field strength but is not well constrained. \added{The explosion time, $t_{\rm exp}$, is the time between explosion and first observation in the pure magnetar model. However, under the assumption that the pre-maximum bumps come from a different power source that we do not model, we can treat it as the time that the magnetar starts powering the light curve, rather than the true supernova explosion time.} The main parameters that constrain the observed properties of a SLSN are the neutron star's initial spin period $P$, magnetic field strength $B$, ejecta mass $M_{\rm ej}$, and ejecta velocity $\varv_{\rm ej}$ (the latter two can be combined to determine the kinetic energy, $E_K$). The model parameters and their priors are listed in Table~\ref{tab:priors}.

For each light curve fit, the first 10,000 iterations are used to burn in the ensemble, during which minimization is employed periodically as the ensemble converges to the global optimum; the remainder of the run-time is used to sample the posterior probability distribution. Convergence is measured by calculating the Gelman–Rubin statistic, or potential scale reduction factor \citep[PSRF;][]{Gelman_1992}, which estimates the extent to which the full parameter space has been explored. We terminate our fits at PSRF $<1.1$, which typically equates to 30,000--60,000 iterations, depending on the the number of data points and the intrinsic scatter around our model.

As noted in \S\ref{sec:data}, and seen in previous SLSNe \citep{Nicholl_Smartt_2016,Nicholl_2017b,Lunnan_2018,Angus_2019}, some of the DES events exhibit pre-peak ``bumps'' in their light curves (Figure~\ref{fig:fits}). These bumps were first recognized by \cite{Leloudas_2012} in the light curve of SN 2006oz and have been interpreted in various ways (\citealt{Piro_2015,Kasen_Metzger_Bildsten_2016,Margalit_2018}). However, since the \texttt{MOSFiT} magnetar engine model does not account for these bumps, we exclude these time ranges from our fits via visual inspection (Figure~\ref{fig:fits}). Whereas \cite{Angus_2019} looked specifically for DES14X3taz-like and DES15S2nr-like bumps in the rest-frame $g$-band light curves, we exclude any unusual detection prior to the main peak in any band. In practice, this only affects a third of the events, and eliminates a small fraction of the data for these events.

\input{priors}

Finally, to place the DES SLSN sample in context, and to investigate any redshift evolution of the physical parameters, we compare our results to a sample of 60 SLSNe from other surveys previously modeled in the same manner \citep{Nicholl_2017b,Villar_2018,Blanchard_2018,Blanchard_2019,Blanchard_2020}; see \citet{Blanchard_2020} for the full list of events and their references. Taken together, the DES and literature SLSN samples span a wide range of properties and redshifts, $z\approx 0.057-1.998$ (see top panel in Figure~\ref{fig:mag_z}).

\section{Light Curve Fits} 
\label{sec:results} 

\begin{figure*}
    \centering
    \includegraphics[width=0.9\textwidth]{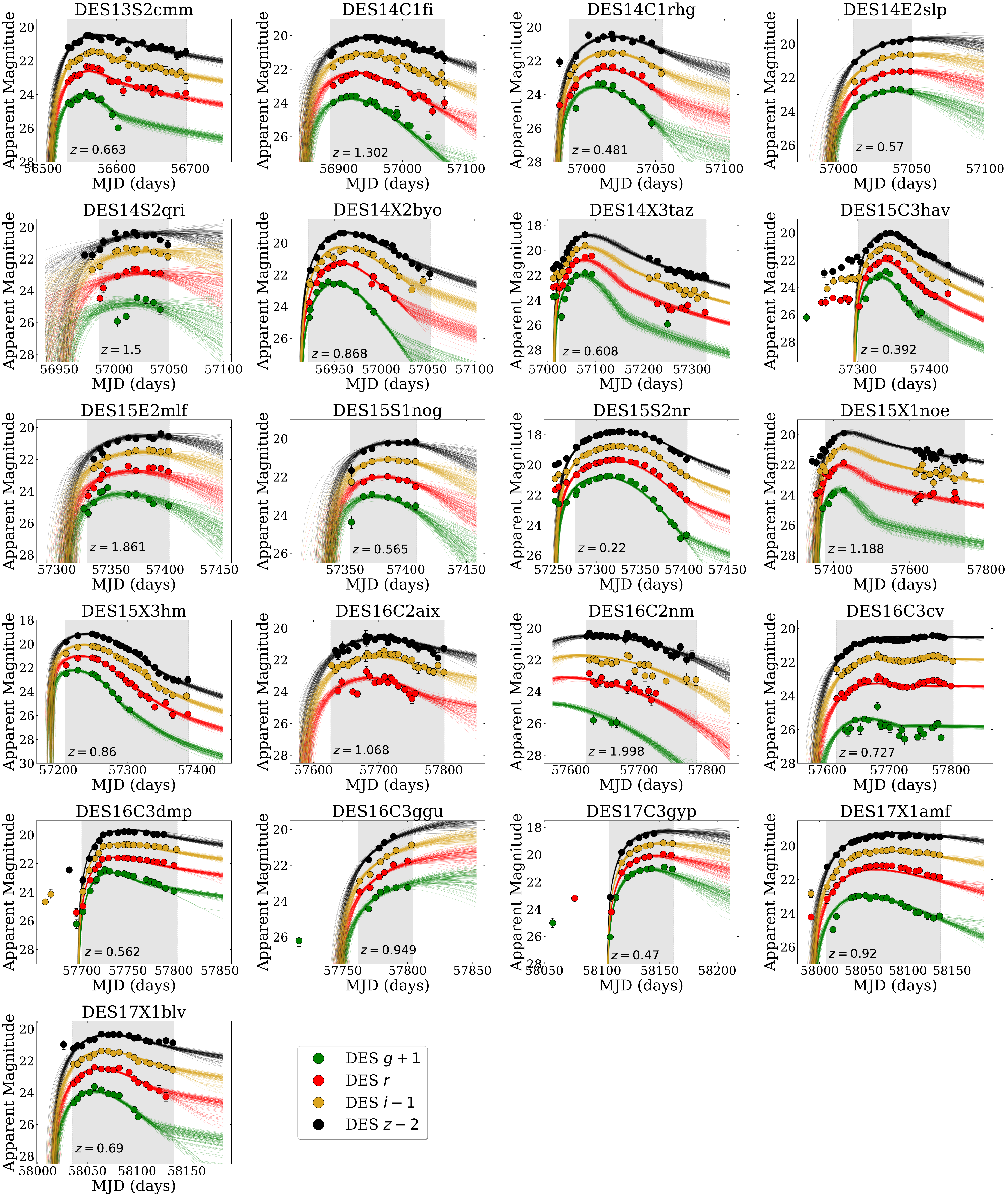}
    \caption{Multi-band extinction-corrected apparent magnitude light curves of the 21 DES SLSNe along with our magnetar model fits using \texttt{MOSFiT}. The different filters are shifted for clarity, as indicated in the legend. The grey shaded regions mark the time ranges used in the fitting, designed to remove significant pre-peak bumps.}
    \label{fig:fits}
\end{figure*}

\input{phys_param}

In Figure~\ref{fig:fits} we show the magnetar model light curve fits. The posterior probability distributions for each parameter consist of the last 5000 steps in each of the 120 Markov chains. We list the parameter median values and $1\sigma$ uncertainties in Table~\ref{tab:phys_params}, where the lower and upper error bars correspond to the 16th and 84th percentiles of the posterior, respectively. We find that most fits are excellent and well converged, and the model parameters are well determined.  The SN colors are well matched, with no apparent systematic offsets. We also include in Table~\ref{tab:phys_params} results from \cite{Angus_2019}. There are several discrepancies in the $P$, $B$, and $M_{\rm ej}$ values outputted by \texttt{MOSFiT} and those reported by \cite{Angus_2019}, which we suspect are caused by different methods used to fit the data (i.e., fitting the pseudo-bolometric light curves vs. multi-band light curves).

Our model includes an intrinsic scatter term, $\sigma$, that when uniformly added to all data points results in a reduced $\chi^2$ of 1; this parameter provides insight into the quality of the fits. We find that the median value for the DES SLSN sample is $\sigma\approx 0.15$ mag, about a factor of 2 higher than the reported photometric errors on individual data points.  This indicates some additional scatter in the data compared to the smooth magnetar models, but an overall reasonable match to the data.  The main discrepancies appear to be small amplitude ``wiggles'' in the light curves that are not captured in the model, and which have been seen in previous SLSNe (e.g., \citealt{Nicholl_2016a}). 

The model posteriors are much narrower than our priors, implying that the priors do not significantly affect our results. The DES SLSN sample median values, and associated $1\sigma$ ranges, of the four key physical parameters, along with the ejecta kinetic energy\footnote{Our model assumes the analytic density profile described in \cite{Margalit_2018}. For a homogeneous density profile, the kinetic energy is given by $E_K=\frac{3}{10}M_{\rm ej}{\varv_{\rm ej}}^2$.}, $E_K=\frac{1}{2}M_{\rm ej}{\varv_{\rm ej}}^2$, are listed in Table~\ref{tab:median}. We also list in the table the values for the SLSN compilation from \citet{Blanchard_2020}. Overall, the model parameters for the DES and literature samples are in good agreement within the $1\sigma$ ranges. In \S\ref{sec:analysis} we examine in more detail the derived parameters, any correlations between them, and any redshift evolution.

\section{Analysis}
\label{sec:analysis}

With the higher redshift range spanned by DES sample, we explore any redshift evolution in the SLSN parameters, or correlations between them. Any redshift evolution would point to evolution in the properties of the progenitors or the magnetar engines over cosmic time, potentially reflecting evolution in the SLSN environments (e.g., metallicity, star formation activity).

\subsection{Bolometric Light Curve Properties}

\input{median}

\begin{figure*}
    \centering
    \includegraphics[width=\columnwidth]{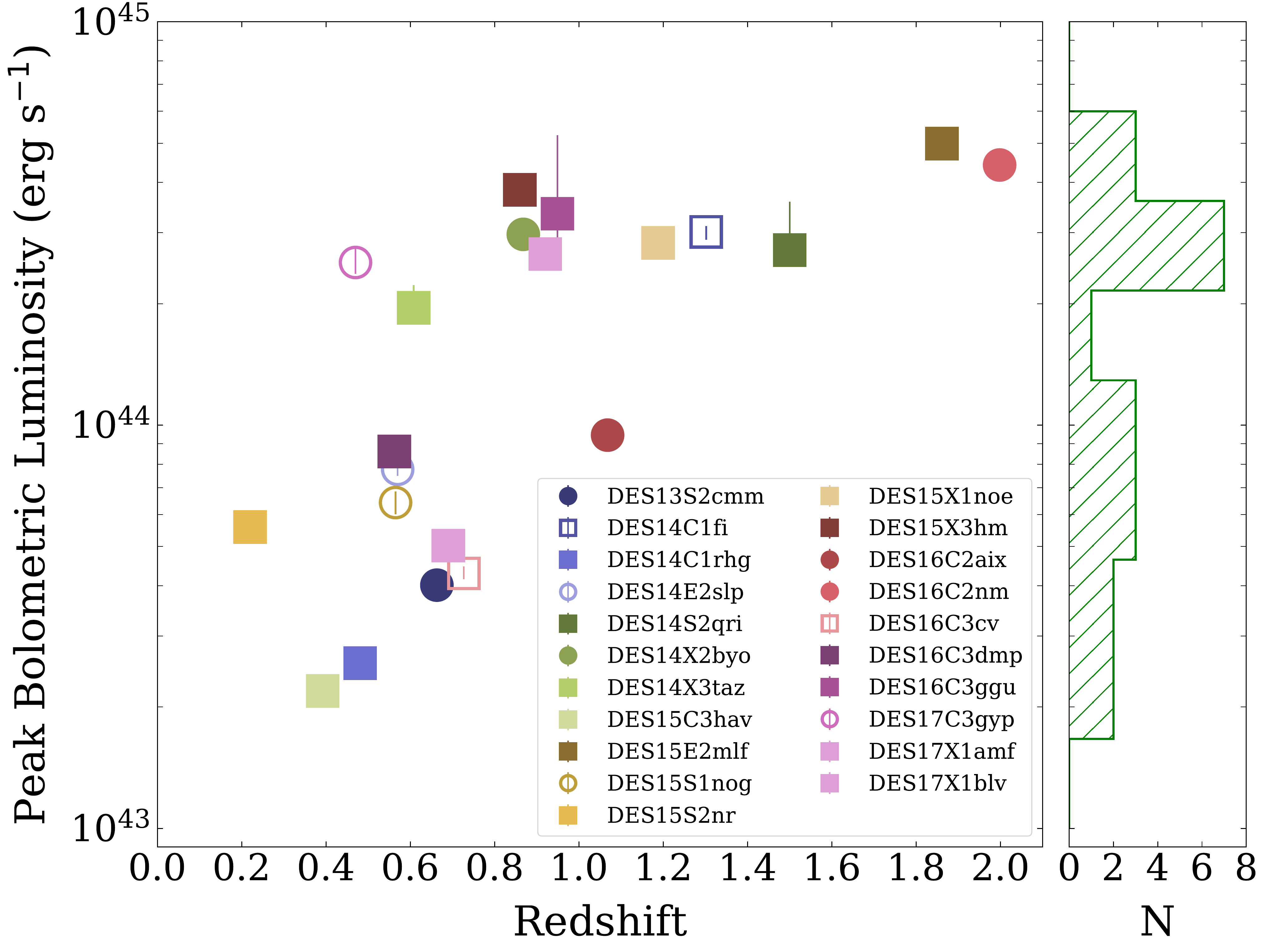}
    \includegraphics[width=\columnwidth]{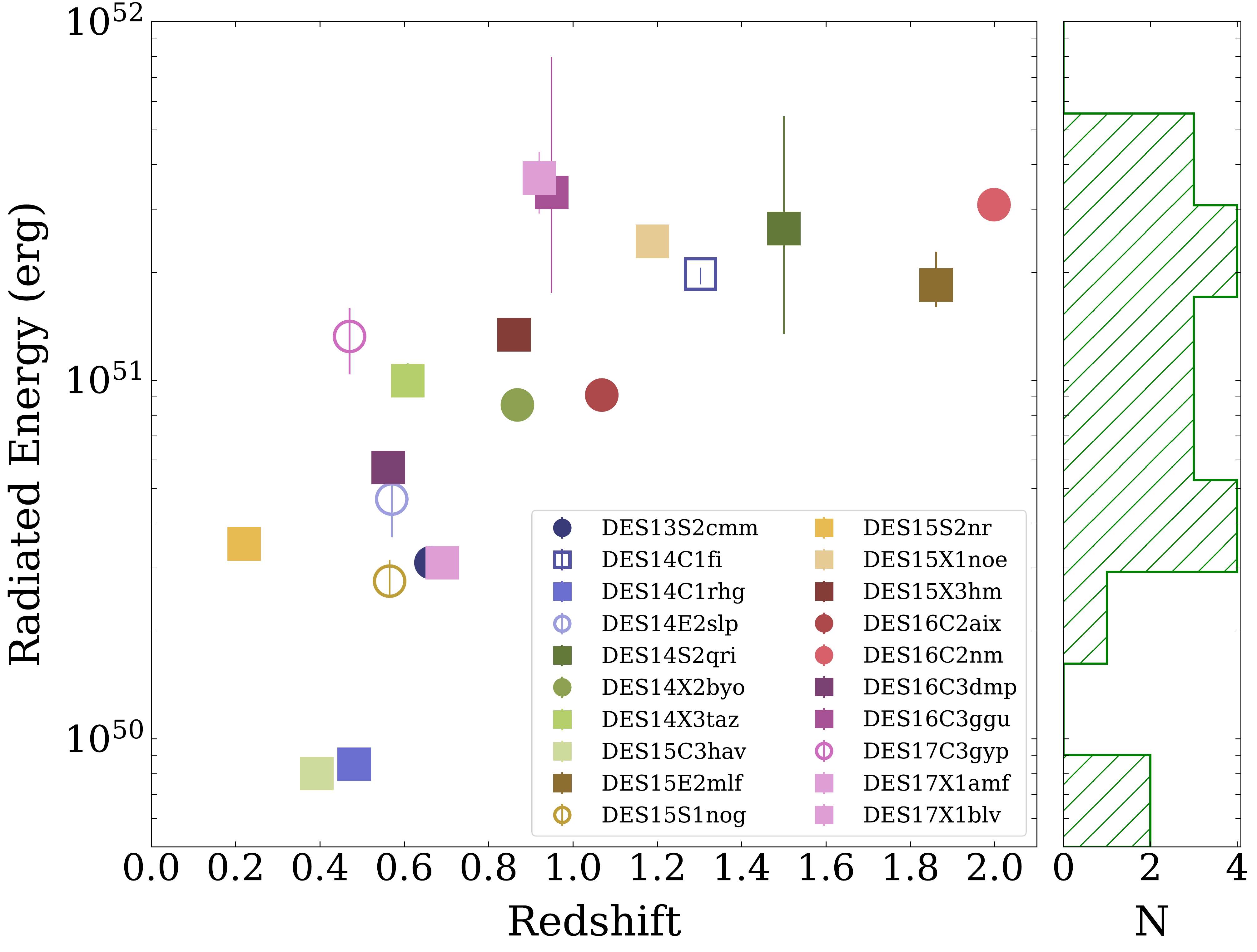}
    \caption{Peak bolometric luminosities (Left) and radiated energies (Right) plotted as functions of redshift.  The values are calculated from the \texttt{MOSFiT} magnetar engine models. We still find evidence for a bimodality in the peak bolometric luminosities, but a less pronounced one in the total radiated energies. The radiated energies plotted here are lower limits, obtained by integrating each bolometric light curve from the explosion time, $t_{\rm exp}$ to 3 $e$-folding times after peak. Square symbols indicate observed pre-peak bumps in the light curve via visual inspection. Solid and open markers indicate the Gold and Silver spectroscopic classification, respectively, of \citet{Angus_2019}.}
    \label{fig:peak_lum}
\end{figure*}

We use the \texttt{MOSFiT} models to generate bolometric light curves for the DES SLSNe, from which we measure the peak luminosities, durations (exponential rise and decline times), and radiated energies. Instead of calculating pseudo-bolometric light curves by summing the observed \textit{griz} photometry, \texttt{MOSFiT} constructs true bolometric light curves before any SED is applied. 

\begin{figure}
    \centering
    \includegraphics[width=\columnwidth]{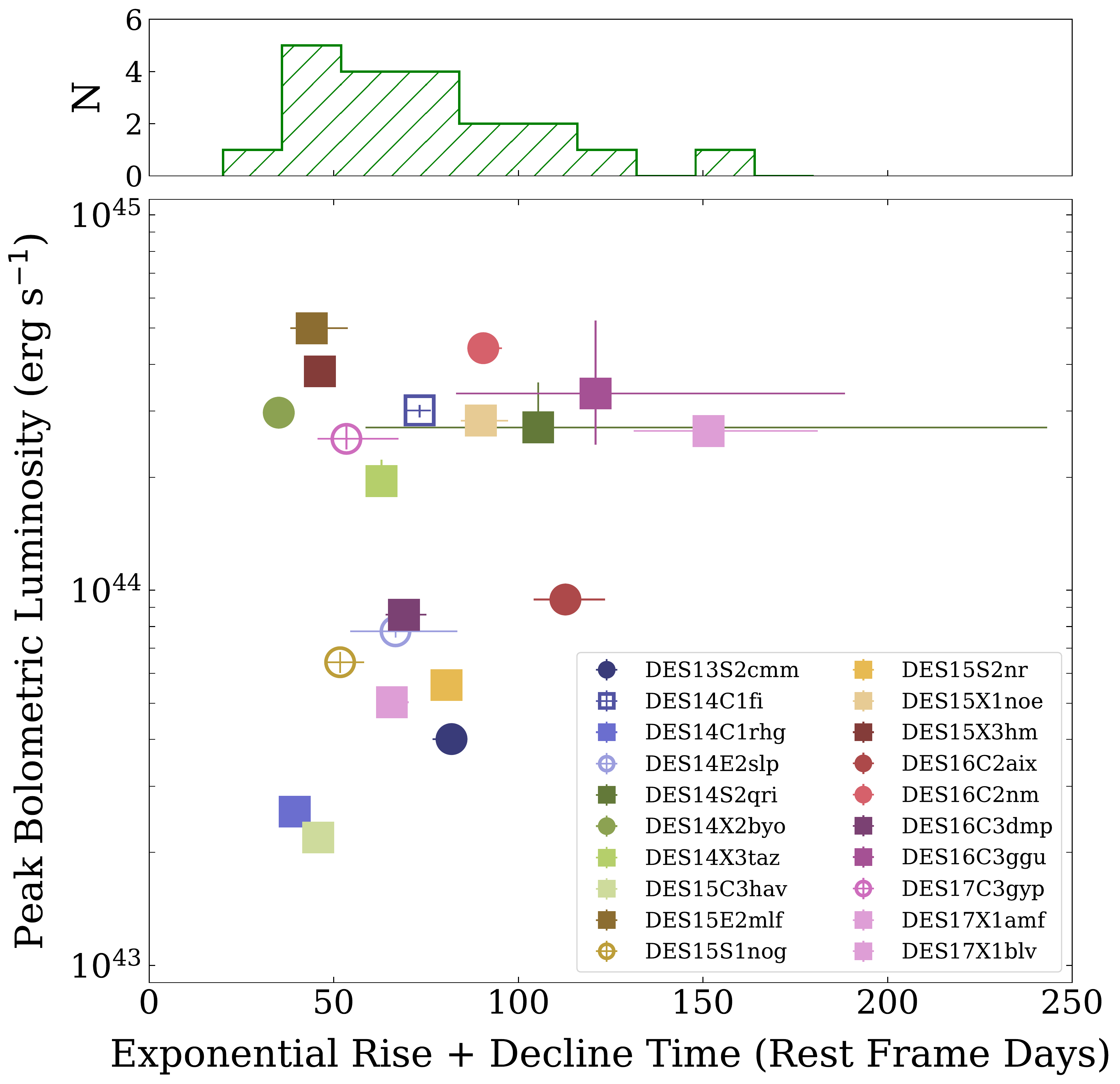}
    \caption{Peak bolometric luminosity versus exponential timescale measured from the model bolometric light curves. Square symbols indicate observed pre-peak bumps in the light curve via visual inspection. Solid and open markers indicate the Gold and Silver spectroscopic classification, respectively, of \citet{Angus_2019}.}
    \label{fig:rise+dec}
\end{figure}

In Figure~\ref{fig:peak_lum} we show the resulting peak bolometric luminosities ($L_{\rm p,bol}$) and radiated energies ($E_{\rm rad}$) versus redshift. In order to compare events with different observed phase ranges, we calculate $E_\mathrm{rad}$ by integrating the bolometric model from the explosion time\added{, $t_{\rm exp}$,} to 3 $e$-folding times after peak, extrapolating the model where necessary. In principle, this should account for the majority of total energy radiated by each SLSN. We find a range of $L_{\rm p,bol}\approx (0.22-4.99)\times 10^{44}$ erg s$^{-1}$, and continued evidence for an apparent bimodal distribution with a gap at $\approx (1-2)\times 10^{44}$ erg s$^{-1}$, as in the case of peak absolute $g$-band magnitudes (\S\ref{sec:data}). For $E_{\rm rad}$, which reflects both the luminosity and the duration of each event, we find a range of $\approx (0.09-3.67)\times10^{51}$ erg and a less pronounced bimodality. This may be due to the distribution of light curve durations, which we show as the combined exponential rise and decline time in Figure~\ref{fig:rise+dec}; we define these as the timescales to brighten from and dim to $L=L_{\rm p,bol}/e$. We find a wide spread in timescale as a function of $L_{\rm p,bol}$, which acts to smooth out the distribution of $E_{\rm rad}$. We exclude DES16C3cv from the right panel of Figure~\ref{fig:peak_lum} and Figure~\ref{fig:rise+dec} because its model light curve, affected by the roughly equal-magnitude peaks in the data, does not decline by a factor of $e$ within a reasonable time frame.

\subsection{Physical Properties and Correlations}

\begin{figure*}
    \centering
    \includegraphics[width=\textwidth]{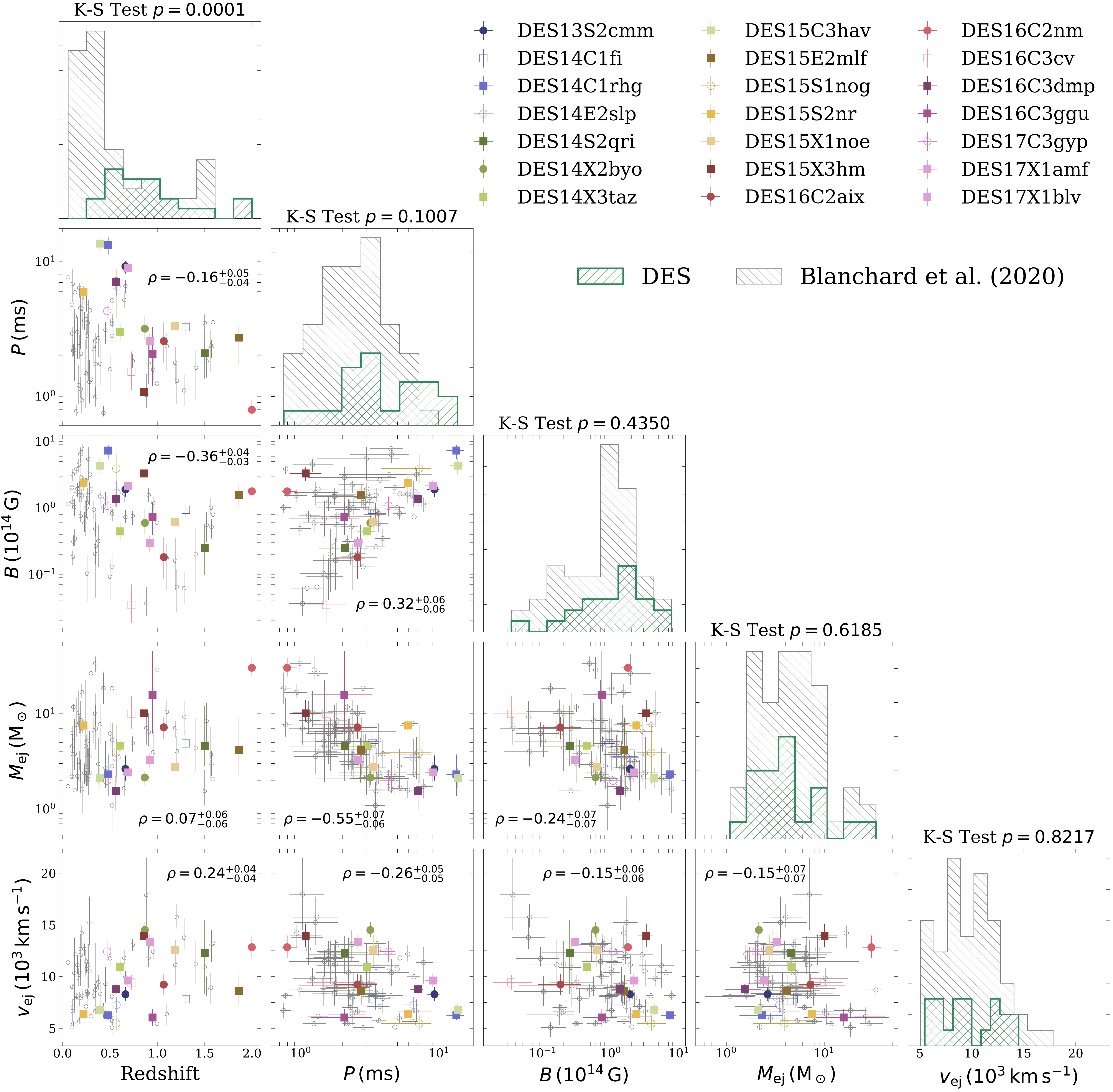}    
    \caption{Median values and 1$\sigma$ uncertainties of the key model parameters ($P$, $B$, $M_{\rm ej}$, $\varv_{\rm ej}$). The gray points mark events from \citet{Blanchard_2020} while the colored points are DES SLSNe modeled in this paper. In the top panels we compare the parameter distributions for the DES (green) and literature (grey) samples, along with the $p$-value associated with the K-S test statistic.  Besides redshift, we conclude that the physical parameters for both samples are drawn from the same distribution. In each panel we quote the median value and 1$\sigma$ bound of the Spearman rank correlation coefficient using the combined DES and literature data set. Of all pairs, $P$ and $M_{\rm ej}$ exhibit the strongest correlation, consistent with the finding in \cite{Blanchard_2020}.}
    \label{fig:phys_params}
\end{figure*}

In Figure~\ref{fig:phys_params} we show two-dimensional distributions of the primary physical parameters ($P$, $B$, $M_{\rm ej}$, and $\varv_{\rm e }$) and the redshifts of the DES SLSNe and the comparison sample from \citet{Blanchard_2020}.  We explore both differences between the two samples, and parameter correlations for the combined sample. Specifically, we compare the DES and literature samples using the two-sample Kolmogorov-Smirnov (K-S) test, and use the resulting $p$-values to determine if both are drawn from the same underlying distribution.  The values are quoted in Figure~\ref{fig:phys_params}.  We find that the only parameter in which the DES and literature samples differ significantly is the redshift. This is due to the nature of the DES survey (deep pencil beam) compared to for example PTF (shallow wide field).  In terms of the magnetar model parameters we find that the distirbutions are in good agreement; the most apparent difference is in $P$ (with $p\approx 0.1$), but this is not a statistically significant result. Thus, we conclude that the DES SLSNe are similar to those found in other surveys, and simply tend to be at somewhat higher redshifts.

With this in mind, we combine the DES and literature samples to explore parameters correlations, including correlations with redshift.  For each pair of parameters, we take uncertainties into account and perform a Monte Carlo procedure to calculate the Spearman rank correlation coefficient \citep[$\rho$;][]{Spearman_1904} and its associated $1\sigma$ bound using the method described in \citet{Curran_2014}. For the model parameters, we find that most combinations exhibit either no correlation (e.g., $M_{\rm ej}$ versus $\varv_{\rm ej}$) or mild correlations that are primarily due to the absence of events in specific subsets of the parameter space. For example, we find $\rho\approx 0.32$ for $P$ versus $B$, which is due primarily to the absence of observed events in the slow spin and low $B$-field portion of the parameter space. This is likely an observational bias, which we investigate in \S\ref{sec:bias}.  The strongest correlation we find is $M_{\rm ej}$ versus $P$ with $\rho\approx -0.55$. This correlation was noted and discussed in \citet{Blanchard_2020}. We see no obvious bimodality in the distribution of any physical parameter.

In terms of redshift evolution, we find that none of the physical parameters appear to be strongly correlated with redshift ($|\rho| \lesssim 0.36$  for all parameters). We therefore conclude that at least to $z\approx 2$ (look-back time of $\approx 10.4$ Gyr) there is no evidence for redshift evolution in the properties of the engine and/or ejecta in SLSNe.

\subsection{The Effect of Observational Biases}
\label{sec:bias}

\begin{figure*}
    \centering
    \includegraphics[width=0.9\textwidth]{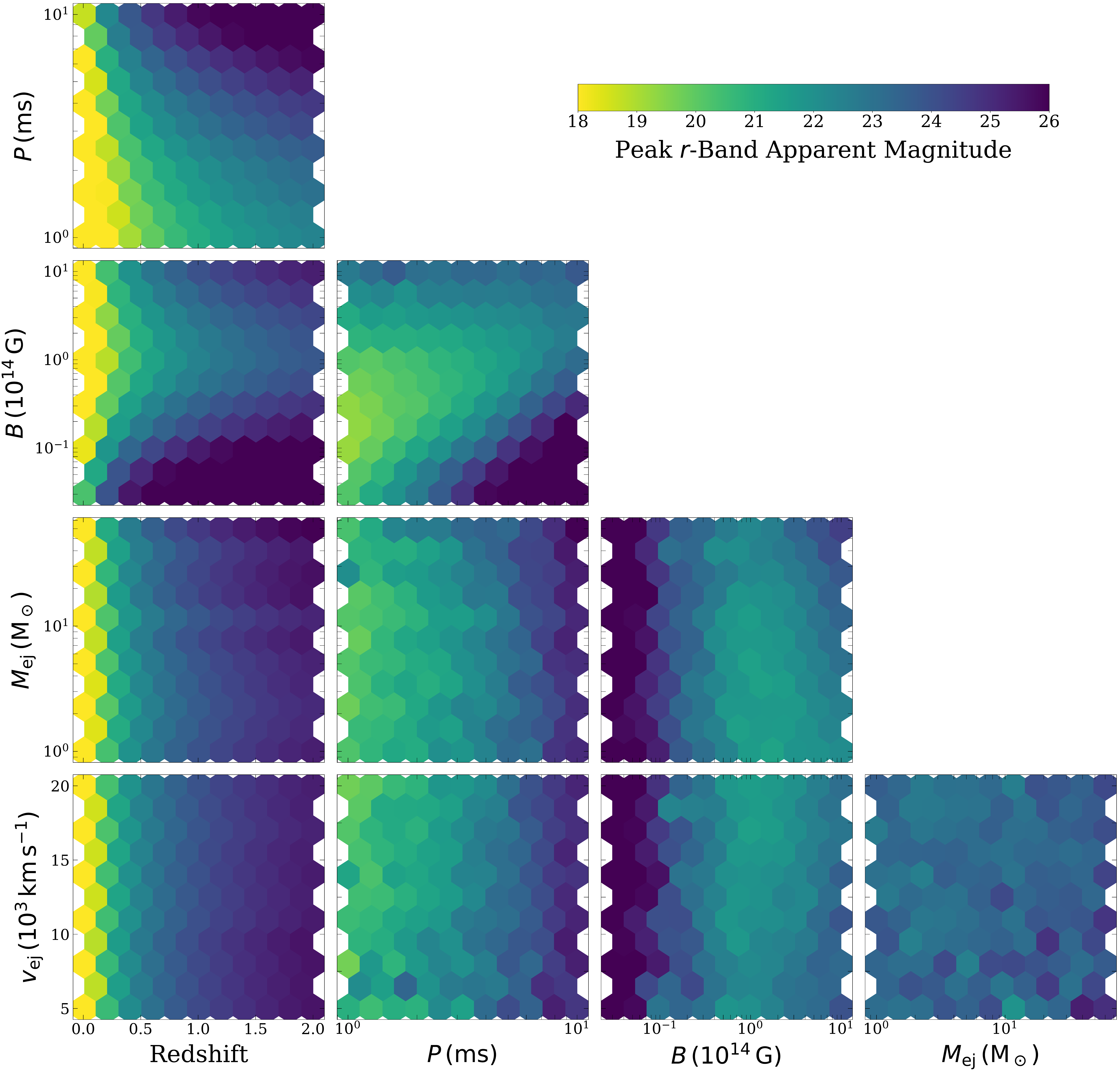}
    \caption{The rest-frame $r$-band apparent magnitude as a function of each pair of physical parameters using 600,000 simulated light curves. The magnitude reported is the sample mean of simulations in each hexagonal cell. The general trend in the $B-P$ space is consistent with observed data, which signifies an observational selection bias. On the other hand, we do not see the strong correlation between $M_{\rm ej}$ and $P$ as suggested by observed data. We can therefore conclude that the $M_{\rm ej}-P$ correlation is a physical effect.}
    \label{fig:grid}
\end{figure*}

Certain combinations of the magnetar engine and/or ejecta parameters lead to fainter light curves that are less easily detectable by observational surveys.  Such an effect might lead to apparent correlations in the model parameters that are driven by observational bias rather than an actual physical effect.  To address this possibility we use a grid of 3,000 simulated SLSN light curves from \citet{Blanchard_2020}, sampled from uniform prior distributions of $P=1–8$ ms, $B=(0.03–10)\times10^{14}$ G, $M_{\rm ej}=1–60\ \rm M_\odot$, and a Gaussian distribution of $\varv_{\rm ej}$ with a mean of $1.47\times10^4$ km s$^{-1}$ and a standard deviation of $4.3\times10^{4}$ km s$^{-1}$ (based on values reported by \citealt{Liu_2017}).

For each simulated light curve, we convert the resulting rest-frame $r$-band absolute magnitudes to apparent magnitudes at $z=0.01–2$ in steps of $\Delta z=0.01$. Each of the resulting 600,000 light curves has a unique set of parameters ($z$, $P$, $B$, $M_{\rm ej}$, and $\varv_{\rm ej}$), which we then treat as an individual SLSN. In reality, the processes in which optical time-domain surveys (including DES) identify SLSNe are  more complicated and often require spectroscopy. We focus here solely on whether we can observe any portion of a light curve theoretically. 

In Figure~\ref{fig:grid}, we show the resulting two-dimensional density plots for each pair of parameters. The plot is structured in the same fashion as Figure~\ref{fig:phys_params} to allow a direct comparison. The $r$-band peak magnitude assigned to each hexagonal cell is the mean of all model SLSNe that reside in the cell. We find that several of the general trends observed in the data are observed in this plot, and can therefore be attributed to observational biases. For example, the dearth of observed SLSNe with slow $P$ at high redshift (top left panel of Figure~\ref{fig:phys_params}) is due to the faintness of such events, with peak apparent magnitudes of $\gtrsim 24$.  Similarly, the absence of events with slow spin and low values of $B$ (Figure~\ref{fig:phys_params}) is again seen as a region of peak apparent magnitudes of $\gtrsim 24$ in Figure~\ref{fig:grid}.

On the other hand, the strong negative correlation between $M_{\rm ej}$ and $P$ cannot be explained by an observational bias.  The upper right corner of this parameter space (high $M_{\rm ej}$ and slow $P$) indeed tends to lead to fainter SLSNe, but the opposite combination (low $M_{\rm ej}$ and fast $P$) leads to bright events that should otherwise be detectable in all surveys.  Thus, as noted by \citet{Blanchard_2020}, the absence of SLSNe with this combination of physical parameters seems to point to a physical origin.

\section{Conclusions} 
\label{sec:conclusions}

We carried out the first systematic modeling of the 21 SLSNe discovered in DES using the magnetar engine model implemented in \texttt{MOSFiT}, building on our previous work of modeling SLSNe from other surveys. While the DES sample exhibits a wide range of light curve luminosities and shapes, the sample can be well explained with the magnetar model.  We find significant overlap between the model parameter distributions of the DES sample and previous SLSNe from various surveys (evaluated with a K-S test), despite the overall higher redshift of the DES sample.  

Examining possible correlations between the model physical parameters ($P$, $B$, $M_{\rm ej}$, and $\varv_{\rm ej}$) for the combined sample of about 80 SLSNe we recover the strong negative correlation between $M_{\rm ej}$ and $P$ previously found by \citet{Blanchard_2020}.  We further find regions devoid of observed SLSNe in some of the two-dimensional parameter spaces.  However, by generating a grid of models across a wide redshift range we find that these regions of parameter space lead to dim events below the detection limit of DES and other surveys.  

Using the combined SLSN sample, and taking advantage of the higher redshifts spanned by the DES (and PS1-MDS; \citealt{Lunnan_2018}) events, we investigate possible redshift evolution of the engine and ejecta parameters. We do not find any clear evidence for such evolution, with no significant correlation ($|\rho|\lesssim0.36$) between redshift and any physical parameter \added{when the modeling is done consistently with \texttt{MOSFiT}}.

Upcoming optical surveys with greater sensitivity and wider field (i.e., Rubin Observatory and Roman Space Telescope) will discover thousands of SLSNe per year to $z\approx 5$ \citep{Villar_2018}. These samples will significantly expand our ability to discern evolution in SLSN properties.

\acknowledgments
We thank Rosanne Di Stefano and K. Azalee Bostroem for constructive and helpful comments on an early version of this draft. B.~H.\ thanks Ashley Villar, Sebastian Gomez, and Matt Nicholl for help running \texttt{MOSFiT}, and Peter Blanchard for providing simulation data. The computations presented in this work were performed on the FASRC Cannon cluster supported by the FAS Division of Science Research Computing Group at Harvard University. The Berger Time Domain group is supported part by NSF and NASA grants.

\noindent \textit{Software}: Astropy \citep{Astropy_2013,Astropy_2018}, extinction \citep{Barbary_2016}, \texttt{MOSFiT} \citep{Guillochon_2018}, Matplotlib \citep{Hunter_2007}, NumPy \citep{Oliphant_2006}, pymccorrelation \citep{Curran_2014,Privon_2020}, pyphot, Scipy \citep{Virtanen_2020}

\noindent \facility{ADS, DES}

\clearpage

\bibliography{Reference}
\bibliographystyle{aasjournal}
\end{document}

%% file: affil.tex
\newcommand{\CfA}{\affiliation{Center for Astrophysics \textbar{} Harvard \& Smithsonian, 60 Garden Street, Cambridge, MA 02138-1516, USA}}

%% file: kcorrection.tex
\begin{deluxetable*}{lcccCccCCc}
\label{tab:kcorrection}
\tablewidth{\columnwidth} 
\caption{Peak Magnitudes of DES SLSNe in the Rest-frame $g$-band}
\tablehead{\colhead{} &
\colhead{} &
\colhead{Milky Way} &
\colhead{$\lambda_\mathrm{eff}$ of} &
\colhead{Closest} &
\colhead{Observed} &
\colhead{Peak Mag.} &
\colhead{} &
\colhead{Rest-frame} & \colhead{}\\[-7pt]
\colhead{SLSN Name} & 
\colhead{Redshift} &
\colhead{Extinction} &
\colhead{Rest-frame}&
\colhead{Observed}&
\colhead{Peak}& \colhead{MJD} &\colhead{$K$ correction} & \colhead{$g$-band Peak} & \colhead{Standard\tablenotemark{$\dagger$}}\\[-7pt]
\colhead{}& \colhead{}&\colhead{$E(B-V)$} & \colhead{$g$-band (\AA)} & \colhead{Band} & \colhead{Mag.} & \colhead{(J2000)} & \colhead{}& \colhead{Abs. Mag.}& \colhead{}}
\startdata
DES13S2cmm & 0.663 & 0.0284 & \phn7934 & i & 22.42 & 56567.18 & -0.56 & -20.08 & Gold\\
DES14C1fi & 1.302 & 0.0091 & 10982 & z & 22.09 & 56956.35 & -1.13 & -21.65 & Silver\\
DES14C1rhg & 0.481 & 0.0096 & \phn7066 & r & 22.30 & 57012.08 & -0.55 & -19.37 & Gold\\
DES14E2slp & 0.57\phn & 0.0060 & \phn7490 & i & 21.60 & 57042.09 & -0.61 & -20.46 & Silver\\
DES14S2qri & 1.50\phn & 0.0276 & 11927 & z & 22.31 & 57018.05 & -1.18 & -21.76 & Gold\\
DES14X2byo & 0.868 & 0.0257 & \phn8912 & z & 21.37 & 56956.21 & -0.84 & -21.57 & Gold\\
DES14X3taz & 0.608 & 0.0220 & \phn7671 & i & 20.60 & 57078.02 & -0.58 & -21.66 & Gold\\
DES15C3hav & 0.392 & 0.0077 & \phn6641 & r & 21.84 & 57339.14 & -0.38 & -19.48 & Gold\\
DES15E2mlf & 1.861 & 0.0086 & 13649 & z & 22.38 & 57396.05 & -1.73 & -21.71 & Gold\\
DES15S1nog & 0.565 & 0.0541 & \phn7466 & i & 22.08 & 57385.08 & -0.53 & -20.03 & Silver\\
DES15S2nr & 0.220 & 0.0293 & \phn5820 & r & 19.66 & 57319.13 & -0.21 & -20.38 & Gold\\
DES15X1noe & 1.188 & 0.0177 & 10439 & z & 21.88 & 57427.06 & -1.03 & -21.70 & Gold\\
DES15X3hm & 0.860 & 0.0237 & \phn8874 & z & 21.17 & 57249.36 & -0.77 & -21.82 & Gold\\
DES16C2aix & 1.068 & 0.0114 & \phn9866 & z & 22.51 & 57680.13 & -0.80 & -21.03 & Gold\\
DES16C2nm & 1.998 & 0.0123 & 14303 & z & 22.31 & 57627.40 & -1.47 & -22.23 & Gold\\
DES16C3cv & 0.727 & 0.0103 & \phn8239 & i & 22.54 & 57681.14 & -0.49 & -20.28 & Silver\\
DES16C3dmp & 0.562 & 0.0068 & \phn7452 & i & 21.66 & 57750.18 & -0.56 & -20.41 & Gold\\
DES16C3ggu & 0.949 & 0.0074 & \phn9298 & z & 22.38 & 57789.10 & -1.15 & -20.49 & Gold\\
DES17C3gyp & 0.47\phn & 0.0073& \phn7013 & r & 19.96 & 58153.08 & -0.54 & -21.66 & Silver\\
DES17X1amf & 0.92\phn & 0.0207 & \phn9160 & z & 21.30 & 58076.13 & -0.77 & -21.87 & Gold\\
DES17X1blv & 0.69\phn & 0.0213 & \phn8063 & i & 22.39 & 58064.06 & -0.61 & -20.17 & Gold\\[+2pt]
\enddata
\tablenotetext{\dagger}{The confidence level in spectroscopic classification designated in \citet{Angus_2019}}
\tablecomments{All magnitudes have been corrected for Milky Way extinction using \cite{Schlafly_Finkbeiner_2011}. Absolute magnitudes are converted using \citetalias{Planck_2016}.}
\end{deluxetable*}

%% file: priors.tex
\begin{deluxetable}{llcccc}
\tablewidth{\columnwidth} 
\caption{Model Parameters and their Associated Priors }
\tablehead{\colhead{Parameter} & 
\colhead{Prior} & 
\colhead{Min} & 
\colhead{Max} & 
\colhead{Mean} & 
\colhead{S.D.} }
\startdata
$P/\rm ms$ & Flat & 0.7 & 10 & $\cdots$ & $\cdots$\\
$B/10^{14}\ \rm G$ & Flat & 0.1 & 10 & $\cdots$ & $\cdots$\\
$M_{\rm ej}$/M$_{\odot}$ & Flat & 0.1 & 100 & $\cdots$ & $\cdots$\\
$\varv_{\rm ej}/10^4\ \rm km\ s^{-1}$ & Gaussian & 0.1 & 3.0 & 1.47 & 4.3\\
$\kappa/\ \rm g\ cm^{-2}$ & Flat  & 0.05 & 0.2 & $\cdots$ & $\cdots$\\
$\kappa_{\gamma}/\ \rm g\ cm^{-2}$ & Log-flat  & 0.01 & 100 & $\cdots$ & $\cdots$\\
$M_{\rm NS}$/M$_{\odot}$ & Flat  & 1.4 & 2.2 & $\cdots$ & $\cdots$\\
$T_{\rm min}/10^3\ \rm K$ & Gaussian & 3.0 & 10.0 & 6.0 & 1.0\\
$n_{\rm H,host}$ & Log-flat  & $10^{16}$ & $10^{23}$ & $\cdots$ & $\cdots$\\
$t_{\rm exp}/\rm days$ & Flat  & $-100$ & 0 & $\cdots$ & $\cdots$\\
$\sigma$ & Log-flat  & 10$^{-3}$ & 100 & $\cdots$ & $\cdots$\\[+2pt]
\enddata
\tablecomments{$P$ is the initial spin period of the magnetar; $B$ is the magnetic field strength; $M_{\rm ej}$ is the ejecta mass; $\varv_{\rm ej}$ is the ejecta velocity; $\kappa$ is the opacity; $\kappa_{\gamma}$ is the gamma-ray opacity; $M_{\rm NS}$ is the neutron star mass; $T_{\rm min}$ is the photospheric temperature floor; $n_{\rm H,host}$ is the hydrogen number density of the host galaxy; $t_{\rm exp}$ characterizes the time of explosion before the first observed data point; $\sigma$ is the additional fractional uncertainty required to yield a reduced $\chi^2$ of 1. For a detailed description of the model see \cite{Nicholl_2017b}.}
\label{tab:priors}
\end{deluxetable}

%% file: phys_param.tex
\begin{deluxetable*}{lcccccccc}
\tablecaption{Median and $1\sigma$ Values for Key Magnetar Engine Model Parameters \label{tab:phys_params}}
\tablehead{\colhead{} & \multicolumn{2}{c}{$P$} & \multicolumn{2}{c}{$B$} & \multicolumn{2}{c}{$M_{\rm ej}$} &
\colhead{$\varv_{\rm ej}$} & \colhead{$\sigma$}\\[-13pt]
\colhead{SLSN Name} & \colhead{} & \colhead{} & \colhead{} & \colhead{} & \colhead{}& \colhead{} & \colhead{} & \colhead{} \\[-12pt]
\colhead{} & 
\multicolumn{2}{c}{(ms)} & 
\multicolumn{2}{c}{(10$^{14}$ G)} &
\multicolumn{2}{c}{(M$_\odot$)} &
\colhead{(10$^3$ km s$^{-1}$)} &
\colhead{(mag)}}

\startdata
DES13S2cmm & $\phn9.25^{+0.80}_{-1.09}$ & & $1.90^{+0.36}_{-0.42}$ & & $\phn2.63^{+1.01}_{-0.64}$ & & $\phn8.29^{+0.67}_{-0.64}$ & $0.13^{+0.02}_{-0.02}$ \\
DES14C1fi & $\phn3.27^{+0.33}_{-0.42}$ & $(3.64)\phn$ & $0.94^{+0.21}_{-0.24}$ & $(3.12)\phn$ & $\phn4.87^{+1.91}_{-1.35}$ & $(17.32)$ & $\phn7.83^{+0.62}_{-0.57}$ & $0.16^{+0.02}_{-0.02}$ \\
DES14C1rhg\tablenotemark{$\ast$} & $13.33^{+1.83}_{-2.03}$ & $(4.95)\phn$ & $7.33^{+1.61}_{-1.86}$ & $(6.03)\phn$ & $\phn2.31^{+1.43}_{-0.95}$ & $(3.19)\phn$ & $\phn6.26^{+0.46}_{-0.44}$ & $0.12^{+0.03}_{-0.02}$ \\
DES14E2slp & $\phn6.53^{+0.91}_{-1.00}$ & & $1.69^{+0.89}_{-0.64}$ & & $\phn3.47^{+1.98}_{-1.29}$ & & $\phn7.26^{+0.78}_{-0.76}$ & $0.11^{+0.03}_{-0.03}$ \\
DES14S2qri\tablenotemark{$\ast$} & $\phn2.09^{+0.96}_{-0.73}$ & $(4.95)\phn$& $0.25^{+0.32}_{-0.15}$&  $(6.03)\phn$& $\phn4.56^{+7.39}_{-1.77}$ & $(3.19)\phn$& $12.30^{+3.20}_{-3.98}$ & $0.27^{+0.06}_{-0.05}$ \\
DES14X2byo & $\phn3.18^{+0.74}_{-0.55}$ & $(5.88)\phn$& $0.59^{+0.36}_{-0.20}$ & $(4.99)\phn$& $\phn2.14^{+0.35}_{-0.34}$ & $(3.19)\phn$& $14.50^{+0.76}_{-0.72}$ & $0.15^{+0.02}_{-0.02}$ \\
DES14X3taz\tablenotemark{$\ast$} & $\phn3.01^{+0.45}_{-0.45}$ &&  $0.44^{+0.15}_{-0.12}$ & & $\phn4.62^{+0.70}_{-0.69}$& & $10.93^{+0.84}_{-0.81}$ & $0.24^{+0.03}_{-0.03}$ \\
DES15C3hav\tablenotemark{$\ast$} & $13.61^{+1.30}_{-1.65}$& & $4.35^{+1.30}_{-1.12}$ & & $\phn2.12^{+0.59}_{-0.45}$& & $\phn6.80^{+0.56}_{-0.50}$ & $0.23^{+0.03}_{-0.02}$ \\
DES15E2mlf\tablenotemark{$\ast$} & $\phn2.73^{+0.61}_{-1.03}$ & $(5.60)\phn$& $1.58^{+0.70}_{-0.54}$ & $(2.68)\phn$& $\phn4.17^{+4.94}_{-1.88}$ & $(1.10)\phn$& $\phn8.62^{+1.51}_{-1.33}$ & $0.26^{+0.05}_{-0.04}$ \\
DES15S1nog & $\phn7.17^{+1.63}_{-3.36}$ & & $3.91^{+2.48}_{-1.50}$ &&  $\phn3.90^{+3.76}_{-2.04}$ & & $\phn5.47^{+0.82}_{-0.69}$ & $0.12^{+0.03}_{-0.03}$ \\
DES15S2nr\tablenotemark{$\ast$} & $\phn5.93^{+0.63}_{-0.74}$ & $(10.52)$& $2.37^{+0.54}_{-0.55}$ & $(12.47)$& $\phn7.52^{+0.96}_{-0.86}$&  $(6.17)\phn$& $\phn6.40^{+0.21}_{-0.21}$ & $0.13^{+0.01}_{-0.01}$ \\
DES15X1noe\tablenotemark{$\ast$} & $\phn3.34^{+0.30}_{-0.40}$ & $(7.45)\phn$& $0.62^{+0.12}_{-0.14}$ & $(2.36)\phn$& $\phn2.75^{+0.90}_{-0.59}$ & $(1.15)\phn$& $12.55^{+1.47}_{-1.59}$ & $0.16^{+0.04}_{-0.04}$ \\
DES15X3hm & $\phn1.08^{+0.33}_{-0.25}$ & $(5.63)\phn$& $3.31^{+0.78}_{-0.79}$ & $(3.49)\phn$& $10.06^{+3.91}_{-1.86}$ & $(4.18)\phn$& $13.93^{+1.20}_{-1.06}$ & $0.14^{+0.02}_{-0.01}$ \\
DES16C2aix & $\phn2.57^{+0.97}_{-0.80}$& & $0.18^{+0.18}_{-0.10}$& & $\phn7.19^{+1.96}_{-1.74}$&&  $\phn9.22^{+1.02}_{-0.99}$ & $0.17^{+0.03}_{-0.03}$ \\
DES16C2nm & $\phn0.80^{+0.15}_{-0.07}$ & & $1.78^{+0.23}_{-0.34}$&&  $30.32^{+7.89}_{-5.76}$ &&  $12.84^{+1.13}_{-1.06}$ & $0.15^{+0.03}_{-0.03}$ \\
DES16C3cv & $\phn1.52^{+0.63}_{-0.41}$& & $0.03^{+0.04}_{-0.02}$&&  $\phn9.95^{+5.37}_{-2.96}$&&  $\phn9.44^{+0.72}_{-0.69}$ & $0.14^{+0.01}_{-0.01}$ \\
DES16C3dmp\tablenotemark{$\ast$} & $\phn7.04^{+0.60}_{-0.88}$ & $(18.27)$& $1.37^{+0.28}_{-0.32}$ & $(9.98)\phn$& $\phn1.54^{+0.71}_{-0.56}$ & $(0.23)\phn$& $\phn8.77^{+0.50}_{-0.49}$ & $0.15^{+0.02}_{-0.01}$ \\
DES16C3ggu\tablenotemark{$\ast$} & $\phn2.06^{+1.21}_{-0.88}$ & $(4.80)\phn$& $0.74^{+0.50}_{-0.35}$ & $(1.05)\phn$& $15.80^{+29.73}_{-9.60}$ & $(16.24)$& $\phn6.05^{+0.82}_{-0.83}$ & $0.14^{+0.03}_{-0.02}$ \\
DES17C3gyp\tablenotemark{$\ast$} & $\phn4.30^{+0.50}_{-0.57}$ & $(6.29)\phn$& $1.07^{+0.41}_{-0.33}$&  $(2.94)\phn$& $\phn1.98^{+0.60}_{-0.44}$ & $(11.57)$& $12.42^{+0.98}_{-1.01}$ & $0.19^{+0.03}_{-0.03}$ \\
DES17X1amf\tablenotemark{$\ast$} & $\phn2.58^{+0.31}_{-0.33}$ & $(5.56)\phn$& $0.30^{+0.09}_{-0.08}$ & $(2.22)\phn$& $\phn3.28^{+0.59}_{-0.52}$ & $(11.39)$& $13.36^{+0.75}_{-0.68}$  & $0.16^{+0.03}_{-0.01}$ \\
DES17X1blv\tablenotemark{$\ast$} & $\phn8.98^{+0.78}_{-1.07}$&&  $2.16^{+0.46}_{-0.50}$&&  $\phn2.42^{+0.68}_{-0.50}$&&  $\phn9.62^{+0.63}_{-0.61}$ & $0.10^{+0.02}_{-0.02}$ \\[+2pt]
\enddata
\tablenotetext{\ast}{SLSNe showing evidence for light curve pre-peak ``bumps''}
\tablecomments{Values quoted in parentheses are the derived physical properties from the magnetar model fits in \cite{Angus_2019}, where available.}
\end{deluxetable*}

%% file: median.tex
\begin{deluxetable}{lcc}
\tablewidth{\columnwidth} 
\caption{DES SLSN Sample Median Values}
\tablehead{\colhead{Parameter} & 
\colhead{DES} &
\colhead{\cite{Blanchard_2020}}}
\startdata
$P$ (ms)& $3.27^{+5.35}_{-1.20}$ & $2.35^{+2.15}_{-1.06}$\\
$B$ ($10^{14}$ G) & $1.37^{+1.75}_{-1.05}$ & $0.98^{+1.17}_{-0.77}$\\
$M_{\rm ej}$ (M$_{\odot}$) & $3.90^{+5.57}_{-1.72}$ & $4.45^{+5.23}_{-2.31}$\\
$\varv_{\rm ej}$ ($10^3$ km s$^{-1}$) & $9.22^{+3.56}_{-2.75}$ & $9.38^{+3.17}_{-2.72}$ \\
$E_K$ ($10^{51}$ erg) & $3.08^{+3.62}_{-1.78}$ & $3.78^{+5.07}_{-1.71}$ \\[+2pt]
\enddata
\tablecomments{The median values and $1\sigma$ ranges for the magnetar engine and ejecta parameters of the DES SLSN sample and the SLSN compilation sample from \citet{Blanchard_2020}.}
\label{tab:median}
\end{deluxetable}

%% file: main.bbl
\begin{thebibliography}{}
\footnotesize
\expandafter\ifx\csname natexlab\endcsname\relax\def\natexlab#1{#1}\fi
\providecommand{\url}[1]{\href{#1}{#1}}
\providecommand{\dodoi}[1]{doi:~\href{http://doi.org/#1}{\nolinkurl{#1}}}

\bibitem[{{Angus} {et~al.}(2019){Angus}, {Smith}, {Sullivan}, {Inserra},
  {Wiseman}, {D'Andrea}, {Thomas}, {Nichol}, {Galbany}, {Childress}, {Asorey},
  {Brown}, {Casas}, {Castander}, {Curtin}, {Frohmaier}, {Glazebrook}, {Gruen},
  {Gutierrez}, {Kessler}, {Kim}, {Lidman}, {Macaulay}, {Nugent}, {Pursiainen},
  {Sako}, {Soares-Santos}, {Thomas}, {Abbott}, {Avila}, {Bertin}, {Brooks},
  {Buckley-Geer}, {Burke}, {Carnero Rosell}, {Carretero}, {da Costa}, {De
  Vicente}, {Desai}, {Diehl}, {Doel}, {Eifler}, {Flaugher}, {Fosalba},
  {Frieman}, {Garc{\'\i}a-Bellido}, {Gruendl}, {Gschwend}, {Hartley},
  {Hollowood}, {Honscheid}, {Hoyle}, {James}, {Kuehn}, {Kuropatkin}, {Lahav},
  {Lima}, {Maia}, {March}, {Marshall}, {Menanteau}, {Miller}, {Miquel}, {Ogand
  o}, {Plazas}, {Romer}, {Sanchez}, {Schindler}, {Schubnell}, {Sobreira},
  {Suchyta}, {Swanson}, {Tarle}, {Thomas}, {Tucker}, \& {DES
  Collaboration}}]{Angus_2019}
{Angus}, C.~R., {Smith}, M., {Sullivan}, M., {et~al.} 2019,
  \hypersetup{urlcolor=magenta}\href{https://dx.doi.org/10.1093/mnras/stz1321}{MNRAS},
  \hypersetup{urlcolor=blue}\href{https://ui.adsabs.harvard.edu/abs/2019MNRAS.487.2215A}{487,
  2215}

\bibitem[{{Astropy Collaboration} {et~al.}(2013){Astropy Collaboration},
  {Robitaille}, {Tollerud}, {Greenfield}, {Droettboom}, {Bray}, {Aldcroft},
  {Davis}, {Ginsburg}, {Price-Whelan}, {Kerzendorf}, {Conley}, {Crighton},
  {Barbary}, {Muna}, {Ferguson}, {Grollier}, {Parikh}, {Nair}, {Unther},
  {Deil}, {Woillez}, {Conseil}, {Kramer}, {Turner}, {Singer}, {Fox}, {Weaver},
  {Zabalza}, {Edwards}, {Azalee Bostroem}, {Burke}, {Casey}, {Crawford},
  {Dencheva}, {Ely}, {Jenness}, {Labrie}, {Lim}, {Pierfederici}, {Pontzen},
  {Ptak}, {Refsdal}, {Servillat}, \& {Streicher}}]{Astropy_2013}
{Astropy Collaboration}, {Robitaille}, T.~P., {Tollerud}, E.~J., {et~al.} 2013,
  \hypersetup{urlcolor=magenta}\href{https://dx.doi.org/10.1051/0004-6361/201322068}{A\&A},
  \hypersetup{urlcolor=blue}\href{https://ui.adsabs.harvard.edu/abs/2013A&A...558A..33A}{558,
  A33}

\bibitem[{{Astropy Collaboration} {et~al.}(2018){Astropy Collaboration},
  {Price-Whelan}, {Sip{\H{o}}cz}, {G{\"u}nther}, {Lim}, {Crawford}, {Conseil},
  {Shupe}, {Craig}, {Dencheva}, {Ginsburg}, {Vand erPlas}, {Bradley},
  {P{\'e}rez-Su{\'a}rez}, {de Val-Borro}, {Aldcroft}, {Cruz}, {Robitaille},
  {Tollerud}, {Ardelean}, {Babej}, {Bach}, {Bachetti}, {Bakanov}, {Bamford},
  {Barentsen}, {Barmby}, {Baumbach}, {Berry}, {Biscani}, {Boquien}, {Bostroem},
  {Bouma}, {Brammer}, {Bray}, {Breytenbach}, {Buddelmeijer}, {Burke},
  {Calderone}, {Cano Rodr{\'\i}guez}, {Cara}, {Cardoso}, {Cheedella}, {Copin},
  {Corrales}, {Crichton}, {D'Avella}, {Deil}, {Depagne}, {Dietrich}, {Donath},
  {Droettboom}, {Earl}, {Erben}, {Fabbro}, {Ferreira}, {Finethy}, {Fox},
  {Garrison}, {Gibbons}, {Goldstein}, {Gommers}, {Greco}, {Greenfield},
  {Groener}, {Grollier}, {Hagen}, {Hirst}, {Homeier}, {Horton}, {Hosseinzadeh},
  {Hu}, {Hunkeler}, {Ivezi{\'c}}, {Jain}, {Jenness}, {Kanarek}, {Kendrew},
  {Kern}, {Kerzendorf}, {Khvalko}, {King}, {Kirkby}, {Kulkarni}, {Kumar},
  {Lee}, {Lenz}, {Littlefair}, {Ma}, {Macleod}, {Mastropietro}, {McCully},
  {Montagnac}, {Morris}, {Mueller}, {Mumford}, {Muna}, {Murphy}, {Nelson},
  {Nguyen}, {Ninan}, {N{\"o}the}, {Ogaz}, {Oh}, {Parejko}, {Parley}, {Pascual},
  {Patil}, {Patil}, {Plunkett}, {Prochaska}, {Rastogi}, {Reddy Janga},
  {Sabater}, {Sakurikar}, {Seifert}, {Sherbert}, {Sherwood-Taylor}, {Shih},
  {Sick}, {Silbiger}, {Singanamalla}, {Singer}, {Sladen}, {Sooley},
  {Sornarajah}, {Streicher}, {Teuben}, {Thomas}, {Tremblay}, {Turner},
  {Terr{\'o}n}, {van Kerkwijk}, {de la Vega}, {Watkins}, {Weaver}, {Whitmore},
  {Woillez}, {Zabalza}, \& {Astropy Contributors}}]{Astropy_2018}
{Astropy Collaboration}, {Price-Whelan}, A.~M., {Sip{\H{o}}cz}, B.~M., {et~al.}
  2018,
  \hypersetup{urlcolor=magenta}\href{https://dx.doi.org/10.3847/1538-3881/aabc4f}{AJ},
  \hypersetup{urlcolor=blue}\href{https://ui.adsabs.harvard.edu/abs/2018AJ....156..123A}{156,
  123}

\bibitem[{Barbary(2016)}]{Barbary_2016}
Barbary, K. 2016,
  \hypersetup{urlcolor=magenta}\href{https://dx.doi.org/10.21105/joss.00058}{JOSS},
  \hypersetup{urlcolor=blue}\href{https://ui.adsabs.harvard.edu/abs/2016JOSS....1...58B}{1,
  58}

\bibitem[{{Bernstein} {et~al.}(2012){Bernstein}, {Kessler}, {Kuhlmann},
  {Biswas}, {Kovacs}, {Aldering}, {Crane}, {D'Andrea}, {Finley}, {Frieman},
  {Hufford}, {Jarvis}, {Kim}, {Marriner}, {Mukherjee}, {Nichol}, {Nugent},
  {Parkinson}, {Reis}, {Sako}, {Spinka}, \& {Sullivan}}]{Berstein_2012}
{Bernstein}, J.~P., {Kessler}, R., {Kuhlmann}, S., {et~al.} 2012,
  \hypersetup{urlcolor=magenta}\href{https://dx.doi.org/10.1088/0004-637X/753/2/152}{ApJ},
  \hypersetup{urlcolor=blue}\href{https://ui.adsabs.harvard.edu/abs/2012ApJ...753..152B}{753,
  152}

\bibitem[{{Blanchard} {et~al.}(2020){Blanchard}, {Berger}, {Nicholl}, \&
  {Villar}}]{Blanchard_2020}
{Blanchard}, P.~K., {Berger}, E., {Nicholl}, M., \& {Villar}, V.~A. 2020,
  \hypersetup{urlcolor=magenta}\href{https://dx.doi.org/10.3847/1538-4357/ab9638}{ApJ},
  \hypersetup{urlcolor=blue}\href{https://ui.adsabs.harvard.edu/abs/2020ApJ...897..114B}{897,
  114}

\bibitem[{{Blanchard} {et~al.}(2019){Blanchard}, {Nicholl}, {Berger},
  {Chornock}, {Milisavljevic}, {Margutti}, \& {Gomez}}]{Blanchard_2019}
{Blanchard}, P.~K., {Nicholl}, M., {Berger}, E., {et~al.} 2019,
  \hypersetup{urlcolor=magenta}\href{https://dx.doi.org/10.3847/1538-4357/aafa13}{ApJ},
  \hypersetup{urlcolor=blue}\href{https://ui.adsabs.harvard.edu/abs/2019ApJ...872...90B}{872,
  90}

\bibitem[{{Blanchard} {et~al.}(2018){Blanchard}, {Nicholl}, {Berger},
  {Chornock}, {Margutti}, {Milisavljevic}, {Fong}, {MacLeod}, \&
  {Bhirombhakdi}}]{Blanchard_2018}
{Blanchard}, P.~K., {Nicholl}, M., {Berger}, E., {et~al.} 2018,
  \hypersetup{urlcolor=magenta}\href{https://dx.doi.org/10.3847/1538-4357/aad8b9}{ApJ},
  \hypersetup{urlcolor=blue}\href{https://ui.adsabs.harvard.edu/abs/2018ApJ...865....9B}{865,
  9}

\bibitem[{Chomiuk {et~al.}(2011)Chomiuk, Chornock, Soderberg, Berger,
  Chevalier, Foley, Huber, Narayan, Rest, Gezari, Kirshner, Riess, Rodney,
  Smartt, Stubbs, Tonry, Wood-Vasey, Burgett, Chambers, Czekala, Flewelling,
  Forster, Kaiser, Kudritzki, Magnier, Martin, Morgan, Neill, Price, Roth,
  Sanders, \& Wainscoat}]{Chomiuk_2011}
Chomiuk, L., Chornock, R., Soderberg, A.~M., {et~al.} 2011,
  \hypersetup{urlcolor=magenta}\href{https://dx.doi.org/10.1088/0004-637x/743/2/114}{ApJ},
  \hypersetup{urlcolor=blue}\href{https://ui.adsabs.harvard.edu/abs/2011ApJ...743..114L}{743,
  114}

\bibitem[{{Curran}(2014)}]{Curran_2014}
{Curran}, P.~A. 2014,
  \hypersetup{urlcolor=magenta}\href{https://arxiv.org/abs/1411.3816}{arXiv}{:}\hypersetup{urlcolor=blue}\href{https://ui.adsabs.harvard.edu/abs/2014arXiv1411.3816C}{1411.3816}

\bibitem[{{Dark Energy Survey Collaboration}(2005)}]{DES_2005}
{Dark Energy Survey Collaboration}. 2005,
  \hypersetup{urlcolor=magenta}\href{https://arxiv.org/abs/astro-ph/0510346}{arXiv}{:}\hypersetup{urlcolor=blue}\href{https://ui.adsabs.harvard.edu/abs/2005astro.ph.10346T}{astro-ph/0510346}

\bibitem[{{Dark Energy Survey Collaboration}(2016)}]{DES_2016}
{Dark Energy Survey Collaboration}. 2016,
  \hypersetup{urlcolor=magenta}\href{https://dx.doi.org/10.1093/mnras/stw641}{MNRAS},
  \hypersetup{urlcolor=blue}\href{https://ui.adsabs.harvard.edu/abs/2016MNRAS.460.1270D}{460,
  1270}

\bibitem[{{De Cia} {et~al.}(2018){De Cia}, {Gal-Yam}, {Rubin}, {Leloudas},
  {Vreeswijk}, {Perley}, {Quimby}, {Yan}, {Sullivan}, {Fl{\"o}rs}, {Sollerman},
  {Bersier}, {Cenko}, {Gal-Yam}, {Maguire}, {Ofek}, {Prentice}, {Schulze},
  {Spyromilio}, {Valenti}, {Arcavi}, {Corsi}, {Howell}, {Mazzali}, {Kasliwal},
  {Taddia}, \& {Yaron}}]{DeCia_2018}
{De Cia}, A., {Gal-Yam}, A., {Rubin}, A., {et~al.} 2018,
  \hypersetup{urlcolor=magenta}\href{https://dx.doi.org/10.3847/1538-4357/aab9b6}{ApJ},
  \hypersetup{urlcolor=blue}\href{https://ui.adsabs.harvard.edu/abs/2018ApJ...860..100D}{860,
  100}

\bibitem[{Dessart {et~al.}(2012)Dessart, Hillier, Waldman, Livne, \&
  Blondin}]{Dessart_2012}
Dessart, L., Hillier, D.~J., Waldman, R., Livne, E., \& Blondin, S. 2012,
  \hypersetup{urlcolor=magenta}\href{https://dx.doi.org/10.1111/j.1745-3933.2012.01329.x}{MNRAS},
  \hypersetup{urlcolor=blue}\href{https://ui.adsabs.harvard.edu/abs/2012MNRAS.426L..76D}{426,
  L76}

\bibitem[{{Flaugher} {et~al.}(2015){Flaugher}, {Diehl}, {Honscheid}, {Abbott},
  {Alvarez}, {Angstadt}, {Annis}, {Antonik}, {Ballester}, {Beaufore},
  {Bernstein}, {Bernstein}, {Bigelow}, {Bonati}, {Boprie}, {Brooks},
  {Buckley-Geer}, {Campa}, {Cardiel-Sas}, {Castand er}, {Castilla}, {Cease},
  {Cela-Ruiz}, {Chappa}, {Chi}, {Cooper}, {da Costa}, {Dede}, {Derylo},
  {DePoy}, {de Vicente}, {Doel}, {Drlica-Wagner}, {Eiting}, {Elliott}, {Emes},
  {Estrada}, {Fausti Neto}, {Finley}, {Flores}, {Frieman}, {Gerdes},
  {Gladders}, {Gregory}, {Gutierrez}, {Hao}, {Holland}, {Holm}, {Huffman},
  {Jackson}, {James}, {Jonas}, {Karcher}, {Karliner}, {Kent}, {Kessler},
  {Kozlovsky}, {Kron}, {Kubik}, {Kuehn}, {Kuhlmann}, {Kuk}, {Lahav}, {Lathrop},
  {Lee}, {Levi}, {Lewis}, {Li}, {Mand richenko}, {Marshall}, {Martinez},
  {Merritt}, {Miquel}, {Mu{\~n}oz}, {Neilsen}, {Nichol}, {Nord}, {Ogando},
  {Olsen}, {Palaio}, {Patton}, {Peoples}, {Plazas}, {Rauch}, {Reil}, {Rheault},
  {Roe}, {Rogers}, {Roodman}, {Sanchez}, {Scarpine}, {Schindler}, {Schmidt},
  {Schmitt}, {Schubnell}, {Schultz}, {Schurter}, {Scott}, {Serrano}, {Shaw},
  {Smith}, {Soares-Santos}, {Stefanik}, {Stuermer}, {Suchyta}, {Sypniewski},
  {Tarle}, {Thaler}, {Tighe}, {Tran}, {Tucker}, {Walker}, {Wang}, {Watson},
  {Weaverdyck}, {Wester}, {Woods}, {Yanny}, \& {DES
  Collaboration}}]{Flaugher_2015}
{Flaugher}, B., {Diehl}, H.~T., {Honscheid}, K., {et~al.} 2015,
  \hypersetup{urlcolor=magenta}\href{https://dx.doi.org/10.1088/0004-6256/150/5/150}{AJ},
  \hypersetup{urlcolor=blue}\href{https://ui.adsabs.harvard.edu/abs/2015AJ....150..150F}{150,
  150}

\bibitem[{{Gal-Yam}(2012)}]{Gal-Yam_2012}
{Gal-Yam}, A. 2012,
  \hypersetup{urlcolor=magenta}\href{https://dx.doi.org/10.1126/science.1203601}{Science},
  \hypersetup{urlcolor=blue}\href{https://ui.adsabs.harvard.edu/abs/2012Sci...337..927G}{337,
  927}

\bibitem[{Gelman \& Rubin(1992)}]{Gelman_1992}
Gelman, A., \& Rubin, D.~B. 1992,
  \hypersetup{urlcolor=magenta}\href{https://dx.doi.org/10.1214/ss/1177011136}{StaSc},
  \hypersetup{urlcolor=blue}\href{https://ui.adsabs.harvard.edu/abs/1992StaSc...7..457G}{7,
  457}

\bibitem[{Guillochon {et~al.}(2018)Guillochon, Nicholl, Villar, Mockler,
  Narayan, Mandel, Berger, \& Williams}]{Guillochon_2018}
Guillochon, J., Nicholl, M., Villar, V.~A., {et~al.} 2018,
  \hypersetup{urlcolor=magenta}\href{https://dx.doi.org/10.3847/1538-4365/aab761}{ApJS},
  \hypersetup{urlcolor=blue}\href{https://ui.adsabs.harvard.edu/abs/2018ApJS..236....6G}{236,
  6}

\bibitem[{{Howell} {et~al.}(2005){Howell}, {Sullivan}, {Perrett}, {Bronder},
  {Hook}, {Astier}, {Aubourg}, {Balam}, {Basa}, {Carlberg}, {Fabbro},
  {Fouchez}, {Guy}, {Lafoux}, {Neill}, {Pain}, {Palanque-Delabrouille},
  {Pritchet}, {Regnault}, {Rich}, {Taillet}, {Knop}, {McMahon}, {Perlmutter},
  \& {Walton}}]{Howell_2005}
{Howell}, D.~A., {Sullivan}, M., {Perrett}, K., {et~al.} 2005,
  \hypersetup{urlcolor=magenta}\href{https://dx.doi.org/10.1086/497119}{ApJ},
  \hypersetup{urlcolor=blue}\href{https://ui.adsabs.harvard.edu/abs/2005ApJ...634.1190H}{634,
  1190}

\bibitem[{Hunter(2007)}]{Hunter_2007}
Hunter, J.~D. 2007,
  \hypersetup{urlcolor=magenta}\href{https://dx.doi.org/10.1109/MCSE.2007.55}{CSE},
  \hypersetup{urlcolor=blue}\href{https://ui.adsabs.harvard.edu/abs/2007CSE.....9...90H}{9,
  90}

\bibitem[{{Inserra} {et~al.}(2017){Inserra}, {Nicholl}, {Chen}, {Jerkstrand},
  {Smartt}, {Kr{\"u}hler}, {Anderson}, {Baltay}, {Della Valle}, {Fraser},
  {Gal-Yam}, {Galbany}, {Kankare}, {Maguire}, {Rabinowitz}, {Smith}, {Valenti},
  \& {Young}}]{Inserra_2017}
{Inserra}, C., {Nicholl}, M., {Chen}, T.~W., {et~al.} 2017,
  \hypersetup{urlcolor=magenta}\href{https://dx.doi.org/10.1093/mnras/stx834}{MNRAS},
  \hypersetup{urlcolor=blue}\href{https://ui.adsabs.harvard.edu/abs/2017MNRAS.468.4642I}{468,
  4642}

\bibitem[{{Jerkstrand} {et~al.}(2017){Jerkstrand}, {Smartt}, {Inserra},
  {Nicholl}, {Chen}, {Kr{\"u}hler}, {Sollerman}, {Taubenberger}, {Gal-Yam},
  {Kankare}, {Maguire}, {Fraser}, {Valenti}, {Sullivan}, {Cartier}, \&
  {Young}}]{Jerkstrand_2017}
{Jerkstrand}, A., {Smartt}, S.~J., {Inserra}, C., {et~al.} 2017,
  \hypersetup{urlcolor=magenta}\href{https://dx.doi.org/10.3847/1538-4357/835/1/13}{ApJ},
  \hypersetup{urlcolor=blue}\href{https://ui.adsabs.harvard.edu/abs/2017ApJ...835...13J}{835,
  13}

\bibitem[{{Kasen} \& {Bildsten}(2010)}]{Kasen_Bildsten_2010}
{Kasen}, D., \& {Bildsten}, L. 2010,
  \hypersetup{urlcolor=magenta}\href{https://dx.doi.org/10.1088/0004-637X/717/1/245}{ApJ},
  \hypersetup{urlcolor=blue}\href{https://ui.adsabs.harvard.edu/abs/2010ApJ...717..245K}{717,
  245}

\bibitem[{{Kasen} {et~al.}(2016){Kasen}, {Metzger}, \&
  {Bildsten}}]{Kasen_Metzger_Bildsten_2016}
{Kasen}, D., {Metzger}, B.~D., \& {Bildsten}, L. 2016,
  \hypersetup{urlcolor=magenta}\href{https://dx.doi.org/10.3847/0004-637X/821/1/36}{ApJ},
  \hypersetup{urlcolor=blue}\href{https://ui.adsabs.harvard.edu/abs/2016ApJ...821...36K}{821,
  36}

\bibitem[{{Kessler} {et~al.}(2015){Kessler}, {Marriner}, {Childress},
  {Covarrubias}, {D'Andrea}, {Finley}, {Fischer}, {Foley}, {Goldstein},
  {Gupta}, {Kuehn}, {Marcha}, {Nichol}, {Papadopoulos}, {Sako}, {Scolnic},
  {Smith}, {Sullivan}, {Wester}, {Yuan}, {Abbott}, {Abdalla}, {Allam},
  {Benoit-L{\'e}vy}, {Bernstein}, {Bertin}, {Brooks}, {Carnero Rosell},
  {Carrasco Kind}, {Castander}, {Crocce}, {da Costa}, {Desai}, {Diehl},
  {Eifler}, {Fausti Neto}, {Flaugher}, {Frieman}, {Gerdes}, {Gruen}, {Gruendl},
  {Honscheid}, {James}, {Kuropatkin}, {Li}, {Maia}, {Marshall}, {Martini},
  {Miller}, {Miquel}, {Nord}, {Ogando}, {Plazas}, {Reil}, {Romer}, {Roodman},
  {Sanchez}, {Sevilla-Noarbe}, {Smith}, {Soares-Santos}, {Sobreira}, {Tarle},
  {Thaler}, {Thomas}, {Tucker}, {Walker}, \& {DES
  Collaboration}}]{Kessler_2015}
{Kessler}, R., {Marriner}, J., {Childress}, M., {et~al.} 2015,
  \hypersetup{urlcolor=magenta}\href{https://dx.doi.org/10.1088/0004-6256/150/6/172}{AJ},
  \hypersetup{urlcolor=blue}\href{https://ui.adsabs.harvard.edu/abs/2015AJ....150..172K}{150,
  172}

\bibitem[{{Leloudas} {et~al.}(2012){Leloudas}, {Chatzopoulos}, {Dilday},
  {Gorosabel}, {Vinko}, {Gallazzi}, {Wheeler}, {Bassett}, {Fischer}, {Frieman},
  {Fynbo}, {Goobar}, {Jel{\'\i}nek}, {Malesani}, {Nichol}, {Nordin},
  {{\"O}stman}, {Sako}, {Schneider}, {Smith}, {Sollerman}, {Stritzinger},
  {Th{\"o}ne}, \& {de Ugarte Postigo}}]{Leloudas_2012}
{Leloudas}, G., {Chatzopoulos}, E., {Dilday}, B., {et~al.} 2012,
  \hypersetup{urlcolor=magenta}\href{https://dx.doi.org/10.1051/0004-6361/201118498}{A\&A},
  \hypersetup{urlcolor=blue}\href{https://ui.adsabs.harvard.edu/abs/2012A&A...541A.129L}{541,
  A129}

\bibitem[{Liu {et~al.}(2017)Liu, Wang, Wang, Dai, Yu, \& Peng}]{Liu_2017}
Liu, L.-D., Wang, S.-Q., Wang, L.-J., {et~al.} 2017,
  \hypersetup{urlcolor=magenta}\href{https://dx.doi.org/10.3847/1538-4357/aa73d9}{ApJ},
  \hypersetup{urlcolor=blue}\href{https://ui.adsabs.harvard.edu/abs/2017ApJ...842...26L}{842,
  26}

\bibitem[{{Liu} {et~al.}(2017){Liu}, {Modjaz}, \& {Bianco}}]{Liu_2017b}
{Liu}, Y.-Q., {Modjaz}, M., \& {Bianco}, F.~B. 2017,
  \hypersetup{urlcolor=magenta}\href{https://dx.doi.org/10.3847/1538-4357/aa7f74}{ApJ},
  \hypersetup{urlcolor=blue}\href{https://ui.adsabs.harvard.edu/abs/2017ApJ...845...85L}{845,
  85}

\bibitem[{{Lunnan} {et~al.}(2013){Lunnan}, {Chornock}, {Berger},
  {Milisavljevic}, {Drout}, {Sanders}, {Challis}, {Czekala}, {Foley}, {Fong},
  {Huber}, {Kirshner}, {Leibler}, {Marion}, {McCrum}, {Narayan}, {Rest},
  {Roth}, {Scolnic}, {Smartt}, {Smith}, {Soderberg}, {Stubbs}, {Tonry},
  {Burgett}, {Chambers}, {Kudritzki}, {Magnier}, \& {Price}}]{Lunnan_2013}
{Lunnan}, R., {Chornock}, R., {Berger}, E., {et~al.} 2013,
  \hypersetup{urlcolor=magenta}\href{https://dx.doi.org/10.1088/0004-637X/771/2/97}{ApJ},
  \hypersetup{urlcolor=blue}\href{https://ui.adsabs.harvard.edu/abs/2013ApJ...771...97L}{771,
  97}

\bibitem[{{Lunnan} {et~al.}(2014){Lunnan}, {Chornock}, {Berger}, {Laskar},
  {Fong}, {Rest}, {Sanders}, {Challis}, {Drout}, {Foley}, {Huber}, {Kirshner},
  {Leibler}, {Marion}, {McCrum}, {Milisavljevic}, {Narayan}, {Scolnic},
  {Smartt}, {Smith}, {Soderberg}, {Tonry}, {Burgett}, {Chambers}, {Flewelling},
  {Hodapp}, {Kaiser}, {Magnier}, {Price}, \& {Wainscoat}}]{Lunnan_2014}
{Lunnan}, R., {Chornock}, R., {Berger}, E., {et~al.} 2014,
  \hypersetup{urlcolor=magenta}\href{https://dx.doi.org/10.1088/0004-637X/787/2/138}{ApJ},
  \hypersetup{urlcolor=blue}\href{https://ui.adsabs.harvard.edu/abs/2014ApJ...787..138L}{787,
  138}

\bibitem[{Lunnan {et~al.}(2018)Lunnan, Chornock, Berger, Jones, Rest, Czekala,
  Dittmann, Drout, Foley, Fong, Kirshner, Laskar, Leibler, Margutti,
  Milisavljevic, Narayan, Pan, Riess, Roth, Sanders, Scolnic, Smartt, Smith,
  Chambers, Draper, Flewelling, Huber, Kaiser, Kudritzki, Magnier, Metcalfe,
  Wainscoat, Waters, \& Willman}]{Lunnan_2018}
Lunnan, R., Chornock, R., Berger, E., {et~al.} 2018,
  \hypersetup{urlcolor=magenta}\href{https://dx.doi.org/10.3847/1538-4357/aa9f1a}{ApJ},
  \hypersetup{urlcolor=blue}\href{https://ui.adsabs.harvard.edu/abs/2018ApJ...852...81R}{852,
  81}

\bibitem[{{Margalit} {et~al.}(2018){Margalit}, {Metzger}, {Thompson},
  {Nicholl}, \& {Sukhbold}}]{Margalit_2018}
{Margalit}, B., {Metzger}, B.~D., {Thompson}, T.~A., {Nicholl}, M., \&
  {Sukhbold}, T. 2018,
  \hypersetup{urlcolor=magenta}\href{https://dx.doi.org/10.1093/mnras/sty013}{\mnras},
  \hypersetup{urlcolor=blue}\href{https://ui.adsabs.harvard.edu/abs/2018MNRAS.475.2659M}{475,
  2659}

\bibitem[{{Mazzali} {et~al.}(2016){Mazzali}, {Sullivan}, {Pian}, {Greiner}, \&
  {Kann}}]{Mazzali_2016}
{Mazzali}, P.~A., {Sullivan}, M., {Pian}, E., {Greiner}, J., \& {Kann}, D.~A.
  2016,
  \hypersetup{urlcolor=magenta}\href{https://dx.doi.org/10.1093/mnras/stw512}{MNRAS},
  \hypersetup{urlcolor=blue}\href{https://ui.adsabs.harvard.edu/abs/2016MNRAS.458.3455M}{458,
  3455}

\bibitem[{{Metzger} {et~al.}(2015){Metzger}, {Margalit}, {Kasen}, \&
  {Quataert}}]{Metzger_2015}
{Metzger}, B.~D., {Margalit}, B., {Kasen}, D., \& {Quataert}, E. 2015,
  \hypersetup{urlcolor=magenta}\href{https://dx.doi.org/10.1093/mnras/stv2224}{MNRAS},
  \hypersetup{urlcolor=blue}\href{https://ui.adsabs.harvard.edu/abs/2015MNRAS.454.3311M}{454,
  3311}

\bibitem[{{Morganson} {et~al.}(2018){Morganson}, {Gruendl}, {Menanteau},
  {Carrasco Kind}, {Chen}, {Daues}, {Drlica-Wagner}, {Friedel}, {Gower},
  {Johnson}, {Johnson}, {Kessler}, {Paz-Chinch{\'o}n}, {Petravick}, {Pond},
  {Yanny}, {Allam}, {Armstrong}, {Barkhouse}, {Bechtol}, {Benoit-L{\'e}vy},
  {Bernstein}, {Bertin}, {Buckley-Geer}, {Covarrubias}, {Desai}, {Diehl},
  {Goldstein}, {Gruen}, {Li}, {Lin}, {Marriner}, {Mohr}, {Neilsen}, {Ngeow},
  {Paech}, {Rykoff}, {Sako}, {Sevilla-Noarbe}, {Sheldon}, {Sobreira}, {Tucker},
  {Wester}, \& {DES Collaboration}}]{Morganson_2018}
{Morganson}, E., {Gruendl}, R.~A., {Menanteau}, F., {et~al.} 2018,
  \hypersetup{urlcolor=magenta}\href{https://dx.doi.org/10.1088/1538-3873/aab4ef}{PASP},
  \hypersetup{urlcolor=blue}\href{https://ui.adsabs.harvard.edu/abs/2018PASP..130g4501M}{130,
  074501}

\bibitem[{{Nicholl} {et~al.}(2019){Nicholl}, {Berger}, {Blanchard}, {Gomez}, \&
  {Chornock}}]{Nicholl_2019}
{Nicholl}, M., {Berger}, E., {Blanchard}, P.~K., {Gomez}, S., \& {Chornock}, R.
  2019,
  \hypersetup{urlcolor=magenta}\href{https://dx.doi.org/10.3847/1538-4357/aaf470}{ApJ},
  \hypersetup{urlcolor=blue}\href{https://ui.adsabs.harvard.edu/abs/2019ApJ...871..102N}{871,
  102}

\bibitem[{{Nicholl} {et~al.}(2017{\natexlab{\hspace{0pt}a}}){Nicholl},
  {Berger}, {Margutti}, {Blanchard}, {Guillochon}, {Leja}, \&
  {Chornock}}]{Nicholl_2017a}
{Nicholl}, M., {Berger}, E., {Margutti}, R., {et~al.}
  2017{\natexlab{\hspace{0pt}a}},
  \hypersetup{urlcolor=magenta}\href{https://dx.doi.org/10.3847/2041-8213/aa82b1}{ApJL},
  \hypersetup{urlcolor=blue}\href{https://ui.adsabs.harvard.edu/abs/2017ApJ...845L...8N}{845,
  L8}

\bibitem[{{Nicholl} {et~al.}(2017{\natexlab{\hspace{0pt}b}}){Nicholl},
  {Berger}, {Margutti}, {Blanchard}, {Milisavljevic}, {Challis}, {Metzger}, \&
  {Chornock}}]{Nicholl_2017c}
{Nicholl}, M., {Berger}, E., {Margutti}, R., {et~al.}
  2017{\natexlab{\hspace{0pt}b}},
  \hypersetup{urlcolor=magenta}\href{https://dx.doi.org/10.3847/2041-8213/aa56c5}{ApJL},
  \hypersetup{urlcolor=blue}\href{https://ui.adsabs.harvard.edu/abs/2017ApJ...835L...8N}{835,
  L8}

\bibitem[{{Nicholl} {et~al.}(2017{\natexlab{\hspace{0pt}c}}){Nicholl},
  {Guillochon}, \& {Berger}}]{Nicholl_2017b}
{Nicholl}, M., {Guillochon}, J., \& {Berger}, E.
  2017{\natexlab{\hspace{0pt}c}},
  \hypersetup{urlcolor=magenta}\href{https://dx.doi.org/10.3847/1538-4357/aa9334}{ApJ},
  850, 55

\bibitem[{{Nicholl} \& {Smartt}(2016)}]{Nicholl_Smartt_2016}
{Nicholl}, M., \& {Smartt}, S.~J. 2016,
  \hypersetup{urlcolor=magenta}\href{https://dx.doi.org/10.1093/mnrasl/slv210}{MNRAS},
  \hypersetup{urlcolor=blue}\href{https://ui.adsabs.harvard.edu/abs/2016MNRAS.457L..79N}{457,
  L79}

\bibitem[{{Nicholl} {et~al.}(2015){Nicholl}, {Smartt}, {Jerkstrand}, {Inserra},
  {Sim}, {Chen}, {Benetti}, {Fraser}, {Gal-Yam}, {Kankare}, {Maguire}, {Smith},
  {Sullivan}, {Valenti}, {Young}, {Baltay}, {Bauer}, {Baumont}, {Bersier},
  {Botticella}, {Childress}, {Dennefeld}, {Della Valle}, {Elias-Rosa},
  {Feindt}, {Galbany}, {Hadjiyska}, {Le Guillou}, {Leloudas}, {Mazzali},
  {McKinnon}, {Polshaw}, {Rabinowitz}, {Rostami}, {Scalzo}, {Schmidt},
  {Schulze}, {Sollerman}, {Taddia}, \& {Yuan}}]{Nicholl_2015b}
{Nicholl}, M., {Smartt}, S.~J., {Jerkstrand}, A., {et~al.} 2015,
  \hypersetup{urlcolor=magenta}\href{https://dx.doi.org/10.1093/mnras/stv1522}{MNRAS},
  \hypersetup{urlcolor=blue}\href{https://ui.adsabs.harvard.edu/abs/2015MNRAS.452.3869N}{452,
  3869}

\bibitem[{{Nicholl} {et~al.}(2016{\natexlab{\hspace{0pt}a}}){Nicholl},
  {Berger}, {Smartt}, {Margutti}, {Kamble}, {Alexander}, {Chen}, {Inserra},
  {Arcavi}, {Blanchard}, {Cartier}, {Chambers}, {Childress}, {Chornock},
  {Cowperthwaite}, {Drout}, {Flewelling}, {Fraser}, {Gal-Yam}, {Galbany},
  {Harmanen}, {Holoien}, {Hosseinzadeh}, {Howell}, {Huber}, {Jerkstrand },
  {Kankare}, {Kochanek}, {Lin}, {Lunnan}, {Magnier}, {Maguire}, {McCully},
  {McDonald}, {Metzger}, {Milisavljevic}, {Mitra}, {Reynolds}, {Saario},
  {Shappee}, {Smith}, {Valenti}, {Villar}, {Waters}, \&
  {Young}}]{Nicholl_2016a}
{Nicholl}, M., {Berger}, E., {Smartt}, S.~J., {et~al.}
  2016{\natexlab{\hspace{0pt}a}},
  \hypersetup{urlcolor=magenta}\href{https://dx.doi.org/10.3847/0004-637X/826/1/39}{ApJ},
  \hypersetup{urlcolor=blue}\href{https://ui.adsabs.harvard.edu/abs/2016ApJ...826...39N}{826,
  39}

\bibitem[{{Nicholl} {et~al.}(2016{\natexlab{\hspace{0pt}b}}){Nicholl},
  {Berger}, {Margutti}, {Chornock}, {Blanchard}, {Jerkstrand}, {Smartt},
  {Arcavi}, {Challis}, {Chambers}, {Chen}, {Cowperthwaite}, {Gal-Yam},
  {Hosseinzadeh}, {Howell}, {Inserra}, {Kankare}, {Magnier}, {Maguire},
  {Mazzali}, {McCully}, {Milisavljevic}, {Smith}, {Taubenberger}, {Valenti},
  {Wainscoat}, {Yaron}, \& {Young}}]{Nicholl_2016b}
{Nicholl}, M., {Berger}, E., {Margutti}, R., {et~al.}
  2016{\natexlab{\hspace{0pt}b}},
  \hypersetup{urlcolor=magenta}\href{https://dx.doi.org/10.3847/2041-8205/828/2/L18}{ApJL},
  \hypersetup{urlcolor=blue}\href{https://ui.adsabs.harvard.edu/abs/2016ApJ...828L..18N}{828,
  L18}

\bibitem[{{Nicholl} {et~al.}(2018){Nicholl}, {Blanchard}, {Berger},
  {Alexander}, {Metzger}, {Bhirombhakdi}, {Chornock}, {Coppejans}, {Gomez},
  {Margalit}, {Margutti}, \& {Terreran}}]{Nicholl_2018}
{Nicholl}, M., {Blanchard}, P.~K., {Berger}, E., {et~al.} 2018,
  \hypersetup{urlcolor=magenta}\href{https://dx.doi.org/10.3847/2041-8213/aae70d}{ApJL},
  \hypersetup{urlcolor=blue}\href{https://ui.adsabs.harvard.edu/abs/2018ApJ...866L..24N}{866,
  L24}

\bibitem[{Oliphant(2006)}]{Oliphant_2006}
Oliphant, T.~E. 2006, A guide to {NumPy} (USA: Trelgol Publishing)

\bibitem[{{Perley} {et~al.}(2016){Perley}, {Quimby}, {Yan}, {Vreeswijk}, {De
  Cia}, {Lunnan}, {Gal-Yam}, {Yaron}, {Filippenko}, {Graham}, {Laher}, \&
  {Nugent}}]{Perley_2016}
{Perley}, D.~A., {Quimby}, R.~M., {Yan}, L., {et~al.} 2016,
  \hypersetup{urlcolor=magenta}\href{https://dx.doi.org/10.3847/0004-637X/830/1/13}{ApJ},
  \hypersetup{urlcolor=blue}\href{https://ui.adsabs.harvard.edu/abs/2016ApJ...830...13P}{830,
  13}

\bibitem[{{Piro}(2015)}]{Piro_2015}
{Piro}, A.~L. 2015,
  \hypersetup{urlcolor=magenta}\href{https://dx.doi.org/10.1088/2041-8205/808/2/L51}{ApJL},
  \hypersetup{urlcolor=blue}\href{https://ui.adsabs.harvard.edu/abs/2015ApJ...808L..51P}{808,
  L51}

\bibitem[{{Planck Collaboration} {et~al.}(2016){Planck Collaboration}, {Ade},
  {Aghanim}, {Arnaud}, {Ashdown}, {Aumont}, {Baccigalupi}, {Banday},
  {Barreiro}, {Bartlett}, {Bartolo}, {Battaner}, {Battye}, {Benabed},
  {Beno{\^\i}t}, {Benoit-L{\'e}vy}, {Bernard}, {Bersanelli}, {Bielewicz},
  {Bock}, {Bonaldi}, {Bonavera}, {Bond}, {Borrill}, {Bouchet}, {Boulanger},
  {Bucher}, {Burigana}, {Butler}, {Calabrese}, {Cardoso}, {Catalano},
  {Challinor}, {Chamballu}, {Chary}, {Chiang}, {Chluba}, {Christensen},
  {Church}, {Clements}, {Colombi}, {Colombo}, {Combet}, {Coulais}, {Crill},
  {Curto}, {Cuttaia}, {Danese}, {Davies}, {Davis}, {de Bernardis}, {de Rosa},
  {de Zotti}, {Delabrouille}, {D{\'e}sert}, {Di Valentino}, {Dickinson},
  {Diego}, {Dolag}, {Dole}, {Donzelli}, {Dor{\'e}}, {Douspis}, {Ducout},
  {Dunkley}, {Dupac}, {Efstathiou}, {Elsner}, {En{\ss}lin}, {Eriksen},
  {Farhang}, {Fergusson}, {Finelli}, {Forni}, {Frailis}, {Fraisse},
  {Franceschi}, {Frejsel}, {Galeotta}, {Galli}, {Ganga}, {Gauthier}, {Gerbino},
  {Ghosh}, {Giard}, {Giraud-H{\'e}raud}, {Giusarma}, {Gjerl{\o}w},
  {Gonz{\'a}lez-Nuevo}, {G{\'o}rski}, {Gratton}, {Gregorio}, {Gruppuso},
  {Gudmundsson}, {Hamann}, {Hansen}, {Hanson}, {Harrison}, {Helou},
  {Henrot-Versill{\'e}}, {Hern{\'a}ndez-Monteagudo}, {Herranz}, {Hildebrandt},
  {Hivon}, {Hobson}, {Holmes}, {Hornstrup}, {Hovest}, {Huang}, {Huffenberger},
  {Hurier}, {Jaffe}, {Jaffe}, {Jones}, {Juvela}, {Keih{\"a}nen}, {Keskitalo},
  {Kisner}, {Kneissl}, {Knoche}, {Knox}, {Kunz}, {Kurki-Suonio}, {Lagache},
  {L{\"a}hteenm{\"a}ki}, {Lamarre}, {Lasenby}, {Lattanzi}, {Lawrence}, {Leahy},
  {Leonardi}, {Lesgourgues}, {Levrier}, {Lewis}, {Liguori}, {Lilje},
  {Linden-V{\o}rnle}, {L{\'o}pez-Caniego}, {Lubin}, {Mac{\'\i}as-P{\'e}rez},
  {Maggio}, {Maino}, {Mandolesi}, {Mangilli}, {Marchini}, {Maris}, {Martin},
  {Martinelli}, {Mart{\'\i}nez-Gonz{\'a}lez}, {Masi}, {Matarrese}, {McGehee},
  {Meinhold}, {Melchiorri}, {Melin}, {Mendes}, {Mennella}, {Migliaccio},
  {Millea}, {Mitra}, {Miville-Desch{\^e}nes}, {Moneti}, {Montier}, {Morgante},
  {Mortlock}, {Moss}, {Munshi}, {Murphy}, {Naselsky}, {Nati}, {Natoli},
  {Netterfield}, {N{\o}rgaard-Nielsen}, {Noviello}, {Novikov}, {Novikov},
  {Oxborrow}, {Paci}, {Pagano}, {Pajot}, {Paladini}, {Paoletti}, {Partridge},
  {Pasian}, {Patanchon}, {Pearson}, {Perdereau}, {Perotto}, {Perrotta},
  {Pettorino}, {Piacentini}, {Piat}, {Pierpaoli}, {Pietrobon}, {Plaszczynski},
  {Pointecouteau}, {Polenta}, {Popa}, {Pratt}, {Pr{\'e}zeau}, {Prunet},
  {Puget}, {Rachen}, {Reach}, {Rebolo}, {Reinecke}, {Remazeilles}, {Renault},
  {Renzi}, {Ristorcelli}, {Rocha}, {Rosset}, {Rossetti}, {Roudier},
  {Rouill{\'e} d'Orfeuil}, {Rowan-Robinson}, {Rubi{\~n}o-Mart{\'\i}n},
  {Rusholme}, {Said}, {Salvatelli}, {Salvati}, {Sandri}, {Santos},
  {Savelainen}, {Savini}, {Scott}, {Seiffert}, {Serra}, {Shellard}, {Spencer},
  {Spinelli}, {Stolyarov}, {Stompor}, {Sudiwala}, {Sunyaev}, {Sutton},
  {Suur-Uski}, {Sygnet}, {Tauber}, {Terenzi}, {Toffolatti}, {Tomasi},
  {Tristram}, {Trombetti}, {Tucci}, {Tuovinen}, {T{\"u}rler}, {Umana},
  {Valenziano}, {Valiviita}, {Van Tent}, {Vielva}, {Villa}, {Wade}, {Wandelt},
  {Wehus}, {White}, {White}, {Wilkinson}, {Yvon}, {Zacchei}, \&
  {Zonca}}]{Planck_2016}
{Planck Collaboration}, {Ade}, P.~A.~R., {Aghanim}, N., {et~al.} 2016,
  \hypersetup{urlcolor=magenta}\href{https://dx.doi.org/10.1051/0004-6361/201525830}{A\&A},
  \hypersetup{urlcolor=blue}\href{https://ui.adsabs.harvard.edu/abs/2016A&A...594A..13P}{594,
  A13}

\bibitem[{{Privon} {et~al.}(2020){Privon}, {Ricci}, {Aalto}, {Viti}, {Armus},
  {D{\'\i}az-Santos}, {Gonz{\'a}lez-Alfonso}, {Iwasawa}, {Jeff}, {Treister},
  {Bauer}, {Evans}, {Garg}, {Herrero-Illana}, {Mazzarella}, {Larson}, {Blecha},
  {Barcos-Mu{\~n}oz}, {Charmandaris}, {Stierwalt}, \&
  {P{\'e}rez-Torres}}]{Privon_2020}
{Privon}, G.~C., {Ricci}, C., {Aalto}, S., {et~al.} 2020,
  \hypersetup{urlcolor=magenta}\href{https://dx.doi.org/10.3847/1538-4357/ab8015}{\apj},
  \hypersetup{urlcolor=blue}\href{https://ui.adsabs.harvard.edu/abs/2020ApJ...893..149P}{893,
  149}

\bibitem[{{Quimby} {et~al.}(2011){Quimby}, {Kulkarni}, {Kasliwal}, {Gal-Yam},
  {Arcavi}, {Sullivan}, {Nugent}, {Thomas}, {Howell}, {Nakar}, {Bildsten},
  {Theissen}, {Law}, {Dekany}, {Rahmer}, {Hale}, {Smith}, {Ofek}, {Zolkower},
  {Velur}, {Walters}, {Henning}, {Bui}, {McKenna}, {Poznanski}, {Cenko}, \&
  {Levitan}}]{Quimby_2011}
{Quimby}, R.~M., {Kulkarni}, S.~R., {Kasliwal}, M.~M., {et~al.} 2011,
  \hypersetup{urlcolor=magenta}\href{https://dx.doi.org/10.1038/nature10095}{Natur},
  \hypersetup{urlcolor=blue}\href{https://ui.adsabs.harvard.edu/abs/2011Natur.474..487Q}{474,
  487}

\bibitem[{{Quimby} {et~al.}(2018){Quimby}, {De Cia}, {Gal-Yam}, {Leloudas},
  {Lunnan}, {Perley}, {Vreeswijk}, {Yan}, {Bloom}, {Cenko}, {Cooke}, {Ellis},
  {Filippenko}, {Kasliwal}, {Kleiser}, {Kulkarni}, {Matheson}, {Nugent}, {Pan},
  {Silverman}, {Sternberg}, {Sullivan}, \& {Yaron}}]{Quimby_2018}
{Quimby}, R.~M., {De Cia}, A., {Gal-Yam}, A., {et~al.} 2018,
  \hypersetup{urlcolor=magenta}\href{https://dx.doi.org/10.3847/1538-4357/aaac2f}{ApJ},
  \hypersetup{urlcolor=blue}\href{https://ui.adsabs.harvard.edu/abs/2018ApJ...855....2Q}{855,
  2}

\bibitem[{{Schlafly} \& {Finkbeiner}(2011)}]{Schlafly_Finkbeiner_2011}
{Schlafly}, E.~F., \& {Finkbeiner}, D.~P. 2011,
  \hypersetup{urlcolor=magenta}\href{https://dx.doi.org/10.1088/0004-637X/737/2/103}{ApJ},
  \hypersetup{urlcolor=blue}\href{https://ui.adsabs.harvard.edu/abs/2011ApJ...737..103S}{737,
  103}

\bibitem[{{Schulze} {et~al.}(2018){Schulze}, {Kr{\"u}hler}, {Leloudas},
  {Gorosabel}, {Mehner}, {Buchner}, {Kim}, {Ibar}, {Amor{\'\i}n},
  {Herrero-Illana}, {Anderson}, {Bauer}, {Christensen}, {de Pasquale}, {de
  Ugarte Postigo}, {Gallazzi}, {Hjorth}, {Morrell}, {Malesani}, {Sparre},
  {Stalder}, {Stark}, {Th{\"o}ne}, \& {Wheeler}}]{Schulze_2018}
{Schulze}, S., {Kr{\"u}hler}, T., {Leloudas}, G., {et~al.} 2018,
  \hypersetup{urlcolor=magenta}\href{https://dx.doi.org/10.1093/mnras/stx2352}{\mnras},
  \hypersetup{urlcolor=blue}\href{https://ui.adsabs.harvard.edu/abs/2018MNRAS.473.1258S}{473,
  1258}

\bibitem[{Spearman(1904)}]{Spearman_1904}
Spearman, C. 1904, The American Journal of Psychology, 15, 72

\bibitem[{{Villar} {et~al.}(2018){Villar}, {Nicholl}, \&
  {Berger}}]{Villar_2018}
{Villar}, V.~A., {Nicholl}, M., \& {Berger}, E. 2018,
  \hypersetup{urlcolor=magenta}\href{https://dx.doi.org/10.3847/1538-4357/aaee6a}{ApJ},
  \hypersetup{urlcolor=blue}\href{https://ui.adsabs.harvard.edu/abs/2018ApJ...869..166V}{869,
  166}

\bibitem[{{Virtanen} {et~al.}(2020){Virtanen}, {Gommers}, {Oliphant},
  {Haberland}, {Reddy}, {Cournapeau}, {Burovski}, {Peterson}, {Weckesser},
  {Bright}, {van der Walt}, {Brett}, {Wilson}, {Millman}, {Mayorov}, {Nelson},
  {Jones}, {Kern}, {Larson}, {Carey}, {Polat}, {Feng}, {Moore}, {Vand erPlas},
  {Laxalde}, {Perktold}, {Cimrman}, {Henriksen}, {Quintero}, {Harris},
  {Archibald}, {Ribeiro}, {Pedregosa}, {van Mulbregt}, \& {SciPy 1. 0
  Contributors}}]{Virtanen_2020}
{Virtanen}, P., {Gommers}, R., {Oliphant}, T.~E., {et~al.} 2020,
  \hypersetup{urlcolor=magenta}\href{https://dx.doi.org/10.1038/s41592-019-0686-2}{NatMe},
  \hypersetup{urlcolor=blue}\href{https://ui.adsabs.harvard.edu/abs/2020NatMe..17..261V}{17,
  261}

\bibitem[{{Woosley}(2010)}]{Woosley_2010}
{Woosley}, S.~E. 2010,
  \hypersetup{urlcolor=magenta}\href{https://dx.doi.org/10.1088/2041-8205/719/2/L204}{ApJL},
  \hypersetup{urlcolor=blue}\href{https://ui.adsabs.harvard.edu/abs/2010ApJ...719L.204W}{719,
  L204}

\bibitem[{{Yan} {et~al.}(2017){Yan}, {Quimby}, {Gal-Yam}, {Brown},
  {Blagorodnova}, {Ofek}, {Lunnan}, {Cooke}, {Cenko}, {Jencson}, \&
  {Kasliwal}}]{Yan_2017}
{Yan}, L., {Quimby}, R., {Gal-Yam}, A., {et~al.} 2017,
  \hypersetup{urlcolor=magenta}\href{https://dx.doi.org/10.3847/1538-4357/aa6b02}{ApJ},
  \hypersetup{urlcolor=blue}\href{https://ui.adsabs.harvard.edu/abs/2017ApJ...840...57Y}{840,
  57}

\end{thebibliography}
